\documentclass[conference]{IEEEtran}
\IEEEoverridecommandlockouts
\usepackage{cite}
\usepackage{amsmath,amssymb,amsfonts}
\usepackage{algorithmic}
\usepackage{graphicx}
\usepackage{textcomp}
\usepackage{xcolor}

\def\BibTeX{{\rm B\kern-.05em{\sc i\kern-.025em b}\kern-.08em
    T\kern-.1667em\lower.7ex\hbox{E}\kern-.125emX}}

\usepackage{subcaption}  
\usepackage{array}
\usepackage{multirow} 
\usepackage{multicol}

\usepackage{amsmath}  
\usepackage{listings}
\usepackage[hidelinks]{hyperref}
\usepackage{booktabs} 
\usepackage[shortlabels]{enumitem} 
\usepackage[T1]{fontenc}
\usepackage[scaled=0.85]{beramono}

\usepackage[linesnumbered,ruled,vlined]{algorithm2e}
\usepackage{float}
\usepackage{mathtools}

\SetKwInput{KwInput}{Input}                
\SetKwInput{KwOutput}{Output}              

\def\systemName{\texttt{CollabIoT}\xspace}

\usepackage{etoolbox}
\usepackage[skins]{tcolorbox} 

\usepackage{xcolor}
\definecolor{jsonKeyColor}{rgb}{0.5,0.1,0.1}
\definecolor{jsonStringColor}{rgb}{0.1,0.1,0.7}
\definecolor{jsonNumberColor}{rgb}{0.6,0.2,0.5}
\definecolor{jsonCommentColor}{rgb}{0.3,0.5,0.3}

\lstdefinelanguage{json}{
    basicstyle=\footnotesize\ttfamily,
    commentstyle=\color{jsonCommentColor}, 
    stringstyle=\color{jsonStringColor},   
    numberstyle=\scriptsize\color{gray},
    stepnumber=1,
    numbersep=8pt,
    showstringspaces=false,
    breaklines=true,
    string=[s]{"}{"},
    comment=[l]{:\ "},
    morecomment=[l]{:"},
    literate=
        *{0}{{{\color{jsonNumberColor}0}}}{1}
         {1}{{{\color{jsonNumberColor}1}}}{1}
         {2}{{{\color{jsonNumberColor}2}}}{1}
         {3}{{{\color{jsonNumberColor}3}}}{1}
         {4}{{{\color{jsonNumberColor}4}}}{1}
         {5}{{{\color{jsonNumberColor}5}}}{1}
         {6}{{{\color{jsonNumberColor}6}}}{1}
         {7}{{{\color{jsonNumberColor}7}}}{1}
         {8}{{{\color{jsonNumberColor}8}}}{1}
         {9}{{{\color{jsonNumberColor}9}}}{1}
}

\definecolor{codegreen}{rgb}{0,0.6,0}
\definecolor{codegray}{rgb}{0.5,0.5,0.5}
\definecolor{codepurple}{rgb}{0.58,0,0.82}
\definecolor{backcolour}{rgb}{0.95,0.95,0.92}

\lstdefinestyle{yaml}{
    basicstyle=\color{blue}\linespread{1}\footnotesize,
     rulecolor=\color{black},
     string=[s]{'}{'},
     stringstyle=\color{blue},
     comment=[l]{:},
     commentstyle=\color{black},
     morecomment=[l]{-},
     frame=single
 }
 \lstdefinestyle{mypython}{
    language=Python,
    basicstyle=\ttfamily\small,
    keywordstyle=\color{blue},
    stringstyle=\color{orange},
    commentstyle=\color{gray},
    showstringspaces=false,
    frame=single,
    breaklines=true
}

\usepackage[frozencache,cachedir=.]{minted}
\usepackage{authblk}
\begin{document}


\title{LLM-Driven Auto Configuration for Transient IoT Device Collaboration}
\author[1]{Hetvi Shastri}
\author[1]{Walid A. Hanafy}
\author[1]{Li Wu}
\author[1]{David Irwin}
\author[2]{Mani Srivastava}
\author[1]{Prashant Shenoy}
\affil[1]{University of Massachusetts Amherst}
\affil[2]{University of California Los Angeles}

\maketitle

\begin{abstract}
Today’s Internet of Things (IoT) has evolved from simple sensing and actuation devices to those with embedded processing and intelligent services, enabling rich collaborations between users and their devices. However, enabling such collaboration becomes challenging when transient devices need to interact with host devices in temporarily visited environments. In such cases, fine-grained access control policies are necessary to ensure secure interactions; however, manually implementing them is often impractical for non-expert users. Moreover, at run-time, the system must automatically configure the devices and enforce such fine-grained access control rules. Additionally, the system must address the heterogeneity of devices. 

In this paper, we present \systemName, a system that enables secure and seamless device collaboration in transient IoT environments. \systemName employs a Large language Model (LLM)-driven approach to convert users' high-level intents to fine-grained access control policies. To support secure and seamless device collaboration, \systemName adopts capability-based access control for authorization and uses lightweight proxies for policy enforcement, providing hardware-independent abstractions.

We implement a prototype of \systemName's policy generation and auto configuration pipelines and evaluate its efficacy on an IoT testbed and in large-scale emulated environments. We show that our LLM-based policy generation pipeline is able to generate functional and correct policies with 100\% accuracy. At runtime, our evaluation shows that our system configures new devices in $\sim$150 ms, and our proxy-based data plane incurs network overheads of up to 2 ms and access control overheads up to 0.3 ms. 

\end{abstract}


\section{Introduction}
\label{sec:introduction}
Recent technological advances have led to rapid growth in Internet of Things (IoT) devices in varied domains, such as smart homes, mobile health, and entertainment~\cite{Pirbhulal2017:SecureSmartHome, stojkoska2017:IoTReviewSmartHome, Fortino2015:BodySensorFusion, almotiri2016mobile, santos2016iot, sun2019magichand, Kim2023:Erebus, blanco2020creating, Xu2014:IoT_Industry}. Researchers estimate that the number of deployed IoT devices exceeded 18.8 billion in 2024 and is growing in a near-exponential manner~\cite{Sinha2024:StateofIoT}. Some have even predicted an Internet of Trillion Things where such devices become ubiquitous in our daily lives~\cite{Baird2024:TrendofIoT}. 
Alongside this growth, IoT devices have evolved from passive sensing and actuating devices to smart collaborating devices that interact with one another. For example, motion sensors may trigger smart lights or security cameras upon detecting motion~\cite{Chaudhari2024:Occupancy}.
To support such collaboration,  devices need to interact with one another and share access to their data, connectivity, and resources with other devices and their users~\cite{Lopez2015:EdgeCentric, krishnamurthi2020:overview}. In addition, newer IoT devices are often mobile or nomadic in nature, frequently moving into new environments beyond their home networks.  

Device collaboration in such dynamic and ad hoc settings requires device owners to specify access control policies for visitors, while also requiring visiting devices to be configured to access network resources and IoT services in remote environments.
For example, food delivery robots are now common in many university campuses~\cite{ucla_robot_delivery}, and online retailers are increasingly experimenting with autonomous robots or drones to deliver packages~\cite{Dominos_Delivery}.
In the future, such robots could collaborate with home IoT devices to securely unlock entry points—such as a garage door~\cite{amazon_key}—and place packages or food items inside the home. In this scenario, the homeowner needs to configure access control policies that specify which actions delivery robots are allowed or denied. In particular, access to home devices or services needs to be granted in a fine-grained manner such that only an appropriate subset of them becomes accessible to the delivery robots. For security or privacy reasons, such access may need to be limited in different ways---e.g., restricted to the time period of the visit or rate limited to prevent resource hogging or DDoS attacks~\cite{Vishwakarma2020:DDoS}.  When a delivery robot arrives, the IoT system must automatically configure its access rights and facilitate secure and real-time interaction with home devices. While  policy specification and device configuration can be done manually, such an approach does not scale when the number and frequency of transient guest devices entering and leaving an environment increases significantly. 

Enabling secure and seamless device collaboration in transient IoT environments remains challenging in today's edge environments for several reasons. First,  device ecosystems grow increasingly complex and interconnected, \emph{specifying} access policies and \emph{configuring} devices to access resources based on these policies
becomes highly challenging--especially in transient settings where fine-grained access control is a necessity
~\cite{He2018:Rethinking, Zeng2019:Understanding}. However, manually configuring of policies and devices is both tedious and impractical for non-expert users.  
One approach is to use AI-driven auto configuration to automate these tasks. For example, we can leverage the code generation capabilities of large language models (LLMs) to convert a natural language description of policy intent into structured policies. Such policies can then drive the auto configuration of transient guest devices to provide the desired levels of resource access.
However, the generative nature of LLMs introduces several limitations for our setting. First, their outputs are probabilistic and may contain hallucinated or logically inconsistent information, which poses risks in policy enforcement. Second, LLMs are typically trained on broad, general-purpose datasets and lack exposure to domain-specific corpora, which hinders their ability to generate system-specific access policies. As a result, relying solely on LLMs is insufficient for automatic generation of fine-grain access control policies from high-level specification with the accuracy and reliability required in transient IoT environments.

A second challenge is that enabling seamless device collaboration at runtime presents significant hurdles due to the heterogeneity of devices in edge environments and the scale of possible interactions. IoT devices in edge environments such as buildings or homes are inherently heterogeneous. Devices vary widely in terms of hardware capabilities, operating systems, communication protocols, and runtime environments, and they often expose different vendor-specific APIs. For example, enabling a delivery robot to interact with a smart door lock of a home may require installing specific drivers or SDKs. This setup is not only time-consuming but also not scalable in transient IoT environments where devices arrive and leave frequently. Moreover, as the number of devices increases, the number of possible interactions grows rapidly. Supporting real-time collaboration at this scale requires systems that can dynamically discover devices, determine authorized interactions, and enforce policies efficiently, which remains a significant systems challenge. 

To address these challenges, we present \systemName, a system that enables devices in transient IoT environments to securely collaborate by sharing data, resources, and services in a seamless and secure manner. 
\systemName employs a large language model (LLM)-driven approach to generate fine-grained access control policies from high-level natural language descriptions provided by device owners.  The LLMs produce structured YAML-based policies, which are then validated through a multi-stage pipeline to ensure syntactic correctness, semantic soundness, and alignment with the user's intent and contextual constraints. 

\systemName supports an automated policy-driven configuration pipeline based on capability-based access control tokens, which allows for secure and scalable access control enforcement.  Furthermore, \systemName enables collaboration via lightweight proxies that provide policy enforcement and hardware-independent interfaces.

Lastly, \systemName is designed to be lightweight and can be deployed on edge nodes, such as a home hub or gateway controller, enabling real-time and secure interactions with enhanced privacy.

In designing, implementing, and evaluating \systemName, our paper makes the following contributions: 

\begin{enumerate}[leftmargin=*]
    \item \textbf{LLM-based Policy Generation.} We present an LLM-based policy generation pipeline that generates validated access control policies from natural language descriptions. Our approach includes multi-stage validation to ensure correctness and security. 
    
    \item \textbf{Device Auto Configuration.} Our \systemName system supports auto configuration of transient IoT devices to
    enable secure, seamless, and scalable device collaboration. 
    Specifically, \systemName auto-configures devices with  capability-based access tokens for authorization and utilizes lightweight proxies for policy enforcement, providing hardware-independent abstractions.
    
    \item \textbf{Implementation and Evaluation.} We implement \systemName using Pydantic for structured policy generation, with $\sim$2.6k SLOC. We evaluate \systemName on a real testbed and in large-scale emulations. Our results demonstrate the effectiveness of the LLM-based auto-configuration and granular access control in transient IoT environments. Our results demonstrate that our LLM-based policy generation pipeline is capable of generating functional and accurate policies with 100\% accuracy. At runtime, our evaluation shows that our system configures new devices in $\sim$150ms, and our proxy-based data plane incurs network overheads of up to 2 ms and access control overheads up to 0.3 ms. 
\end{enumerate}

\section{Background and Requirements}
\label{sec:background-related-work}
This section provides background on transient collaborative IoT systems and access control auto-configuration, introduces LLM-based approaches for auto-configuration, and outlines the key system design requirements. 

\subsection{ Transient IoT Devices in Edge Environments}
Our work focuses on IoT devices in edge environments such as a home or a building. We assume that each building or a home is under a different administrative domain, and devices in that network are controlled by the home owner or an administrator. IoT systems in such environments have evolved from simple statically-mounted sensing and actuating devices
to mobile, intelligent devices that interact and collaborate with one another and their environments. For example, many devices such as delivery robots, wearables (e.g., AR glasses), and smartphones are mobile and can interact with other devices. Further, such devices can enter and leave different edge environments due to their mobile nature, making them {\em transient devices} when visiting a different network in a new administrative domain. We refer to devices that are present in their home network environment as {\em native devices}, and while those that  visit a different edge environment become {\em guest devices} in that network. Guest devices need to configured with appropriate credentials (e.g., wifi password or authorization tokens) upon entering a new network to access network and devices resources in that environment. Our hypothesis is that as the scale and frequency of transient device visits grows, such configuration can no longer be done manually (as is done in some networks today).

Some current IoT frameworks already support automated configuration primarily through service discovery, enabling devices to locate and identify each other on a network. Protocols such as UPnP~\cite{10.17487/RFC6970}, Zero-Conf (Apple Bonjour)~\cite{AppleBonjour}, and Jini (Apache River)~\cite{Jini} provide plug-and-play capabilities, allowing devices to advertise and discover services without requiring manual configuration. However, these systems generally assume implicit trust among devices on the same network and offer little to no built-in access control or verification mechanisms. While suitable for small, static environments, these protocols are not suitable for transient devices involving multiple trust levels.

\subsection{Access Control in IoT system}
Since transient environments require multiple trust levels, there is a need for a user-friendly and fine-grained access control system~\cite{He2018:Rethinking,goyal2022securing,Lee2024:FLUID-IOT,Blue2021:Lux}, which was not addressed in current IoT systems. 
For example, smart home applications such as Nest Thermostat~\cite{GoogleNest}, Amazon Alexa~\cite{AmazonAlexa}, and Samsung SmartThings~\cite{smartthings} only support all-or-nothing access for guest users. Application such as Wyze~\cite{Wyze} allows per-device sharing but fail to cover the capability level of sharing. Apple HomeKit~\cite{apple_home} can allow additional users with limited capability options such as full control, view-only control, and local or remote control. Although these approaches offer a user-friendly interface, they often fail to offer fine-grained access control, which is crucial in transient settings, since each device needs its own specific access capabilities.

Enabling fine-grained access control, such as allowing a guest with read-only access to stream from a camera without granting it write access to control  or reconfigure it,  can requires  manual configuration of access control policies. 
In dynamic settings where guest devices frequently enter and leave, manual approaches are cumbersome and do not scale, as they are primarily designed for static and single-user environments. As a result, even though fine-grained access is essential, configuration of these policies at scale is not fully addressed in most user-facing platforms due to its complexity. While access control systems such as AWS IoT~\cite{AWSIOT}, Azure IoT~\cite{AzureIoT}, and Amazon Greengrass~\cite{AWSIoTGreengrass} support detailed, fine-grained policies, they are developer-centric and require technical knowledge with a steep learning curve, which is unsuitable for non-expert users.  

\subsection{LLM-based Policy Generation}
Large language models (LLMs)~\cite{vaswani2017attention} demonstrated a strong performance across various language tasks. 
Trained on massive corpora of data, models like GPT-4~\cite{achiam2023gpt} excel at interpreting high-level descriptions, identifying semantically equivalent expressions, performing multi-step contextual reasoning, and generating coherent natural language~\cite{min2023recent, nam2024using, li2022competition}. These capabilities make LLMs highly valuable for applications that involve generating structured results, such as in programming tasks~\cite{LLMCodeGeneration}.

In particular, the use of LLMs for generating policy configurations has attracted growing interest~\cite{subramaniam2024intent, vatsa2025synthesizing, cheng2025:SayWhatYouMean, yao2024survey}. Their generative power enables the synthesis of configuration policies directly from natural language descriptions, significantly lowering the barrier for non-expert users. However, LLMs have inherent limitations that make them unreliable when used alone, especially in security-sensitive domains. Their outputs are probabilistic and can introduce hallucinated or logically inconsistent content. For example, research work~\cite{vatsa2025synthesizing} explores the use of LLMs to synthesize AWS Identity and Access Management (IAM) policies using zero-shot prompting. The study found that the synthesized policies were frequently incomparable to the ground-truth policies and tended to be overly permissive. These deviations largely stemmed from imprecise or underspecified prompts, highlighting the difficulty of producing accurate and secure policies without structured policy specification mechanisms and additional validations.

\subsection{Design Requirements}
To enable seamless collaboration between guests and native devices and services (e.g., embedded intelligence) running in the host environment, our system should address four design requirements:
\begin{itemize}[leftmargin=0.55cm]
    \item[{\bf \texttt{R1}}] \textbf{Intuitive and Precise Policy Configuration Interface} 
    The system should be able to translate users' high-level intent into machine-readable policies and validate the correctness of the generated policies.
    
    \item[{\bf \texttt{R2}}] \textbf{Granular Access and Rate Control} 
    Given the varying levels of trust in guest users and their devices, it is important to implement granular access and rate control. Different guests may need to interact with different subsets of native devices. Therefore, it is essential to restrict access to certain devices and services, limit the request rates, and define which capabilities are permitted.
     \item[{\bf \texttt{R3}}] \textbf{Auto Configuration} The system should be able to address devices' transient requirements and policy changes by automatically granting access to newly arriving devices, without user intervention.
    
    \item[{\bf \texttt{R4}}] \textbf{Hardware independence} Since different IoT devices expose different vendor-specific APIs, it may not be reasonable to expect a guest to know about all vendor-specific interfaces on native devices (e.g., camera for different vendors). Seamless collaboration requires the ability to use vendor- and hardware-independent abstractions.   
\end{itemize}

\section{\systemName Design}
\label{sec:design}
\begin{figure}[t]
    \centering
    \includegraphics[width=\linewidth]{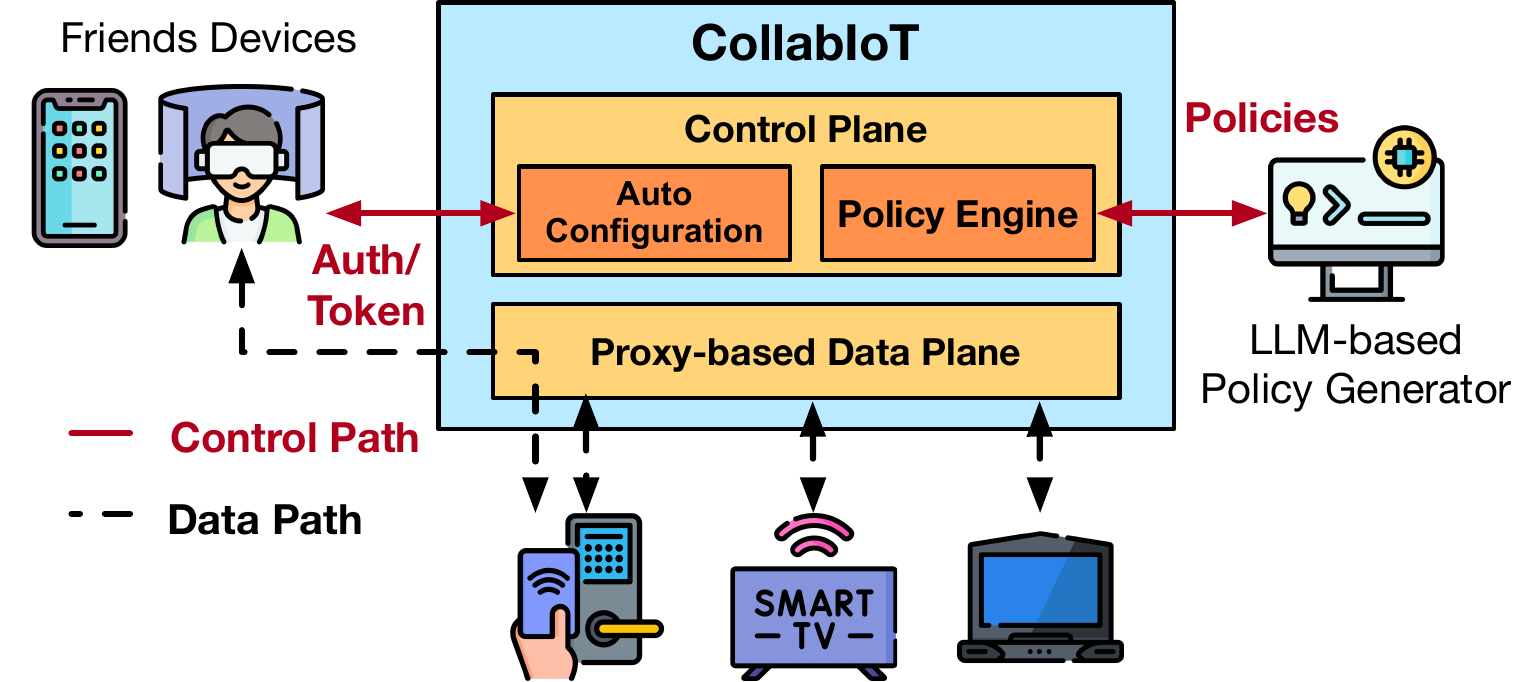}
    \caption{\systemName design overview. 
    }
    \label{fig:overview}
\end{figure}

\begin{table*}[t]
\caption{Examples of hardware-independent attributes, capabilities, and APIs. }
\label{tab:cap_att_examples}
\resizebox{0.95\linewidth}{!}{%

\begin{tabular}{lccc}
        \toprule
        \textbf{Device} & \textbf{Attributes} & \textbf{Capabilities} & \textbf{Generic APIs}\\ \toprule
        \textbf{Phone} & \texttt{owner, sens-level, ID} & NA & NA \\
        \textbf{Bulb} & \texttt{location, type, sens-level, ID, ...} & \texttt{switch, brightness,...} & \texttt{turn\_on(), turn\_off(), set\_brightness()}\\
        
        \textbf{Camera} & \texttt{location, type, sens-level, ID, ...} & \texttt{stream, rotate,...} & \texttt{live\_stream(), rotate(), retrieve()}\\
        
        \textbf{Door lock} & \texttt{location, type, sens-level, ID, ...} & \texttt{switch, configure, logs,...} & \texttt{lock(), unlock(), set\_conf()}\\

        \textbf{Laptop} & \texttt{location, type, sens-level, ID, ...} & \texttt{inference,...} & \texttt{inference\_service()}\\

        \bottomrule
    \end{tabular}
}
\end{table*}

This section presents the architecture and design details of \systemName, our proposed AI-driven auto configuration system for collaborating devices in transient IoT environments. We begin with an overview of our system design, followed by detailed descriptions of the core system components.

\subsection{Design Overview}\label{sec:des_over}
\autoref{fig:overview} depicts the architecture of our \systemName system. At the core of \systemName is a control plane comprising a policy engine and an auto configuration engine that collectively provide automated access to guest devices.
\systemName provides users with an LLM-based policy generation interface, where users specify their intents via natural language, which are compiled into policies. At runtime, when a new device joins the network, the policy engines automatically create an access token based on the device's attributes and system policies.
Lastly, \systemName includes a data plane based on lightweight device proxies that enforce runtime granular access and rate limits and implement hardware-independent interfaces. Next, we provide an overview of \systemName's components:

\noindent\textbf{LLM-based Policy Generator.}
Our LLM-based policy generator enables homeowners or administrators to use text or voice commands, significantly lowering the barrier for non-expert users to manage their IoT environments. Our LLM-based policy generation includes a processing pipeline that compiles the user's prompts into a schema-structured grouping and access control policies.

\noindent\textbf{Policy Engine.}

The policy engine is responsible for validating the correctness of new or LLM-generated policies and maintaining a database of available devices and policies. 

Additionally, when a new device joins the system, the policy engine matches the device attributes with the current system policies. It then generates a list of permissible actions for the specific native devices based on their attributes. 

\noindent\textbf{Auto Configuration Engine.}
The auto configuration engine utilizes the list of permitted actions and creates cryptographically signed capability-based tokens (e.g., a JWT access token) and distributes them to the guest devices.

\noindent\textbf{Proxy-based Data Plane.}
Our proxy-based data plane encapsulates all the device communication via a lightweight proxy that provides a hardware-independent interface and enforces access and rate control on behalf of the device.

In the remainder of this section, we first describe our access control scheme (\autoref{sec:des_spec}), then we describe our LLM-based policy generation pipeline, and then explain our access control auto configuration process (\autoref{sec:auto_config}) and proxy-based communication (\autoref{sec:des_proxy}). Finally, we provide an end-to-end workflow example (\autoref{sec:des_example}).

\subsection{\systemName Access Control Scheme}\label{sec:des_spec}
First, we describe our access control scheme, which comprises a separate dynamic grouping scheme and access control policies that enable the policy engine to construct device groups, map each device to its representative groups, enable resource owners to specify the intended device access policies and determine devices, capabilities, or rate limits across these groups. 

\subsubsection{\textbf{Dynamic Grouping Scheme}}
Our dynamic grouping scheme is an attribute-based method that utilizes device attributes to construct device lists, which simplifies access in dynamic environments. Using attributes to group devices and users is a convenient shorthand over enumerating them, which can be cumbersome. The dynamic grouping policies can use device attributes (e.g., device type or location, such as ``all cameras'' or ``all living room devices''), context attributes (e.g., specify event or times), user attributes (e.g., all my friends), and their combinations to construct groups. We assume each device type $i$, has a set of attributes $A_i$ and capabilities $C_i$ and that all devices (guest or native) use vendor-independent attributes and capabilities descriptions. ~\autoref{tab:cap_att_examples} lists examples of the hardware-independent attributes used by \systemName.

\subsubsection{\textbf{Access Control Policies}}
Our access control policy scheme is a capability-based access control scheme that maps guest and local (native) device groups. The access control policy links different groups by defining source (i.e., guest devices) and destination groups (native devices) using group IDs. 
The access control policies define a per-device-type \texttt{include} and \texttt{exclude} lists that denote allowed and not allowed device capabilities. Additionally, our access control policies feature rate-limiting configurations that enable users to set device limits or priority scheduling policies, further facilitating secure collaboration between devices. Our utilized capabilities combine the hardware-independent capabilities names (see \autoref{tab:cap_att_examples}) within a device type, as capability names may differ across vendors. 
\begin{figure}[t]
    \centering
   \includegraphics[width=\linewidth]{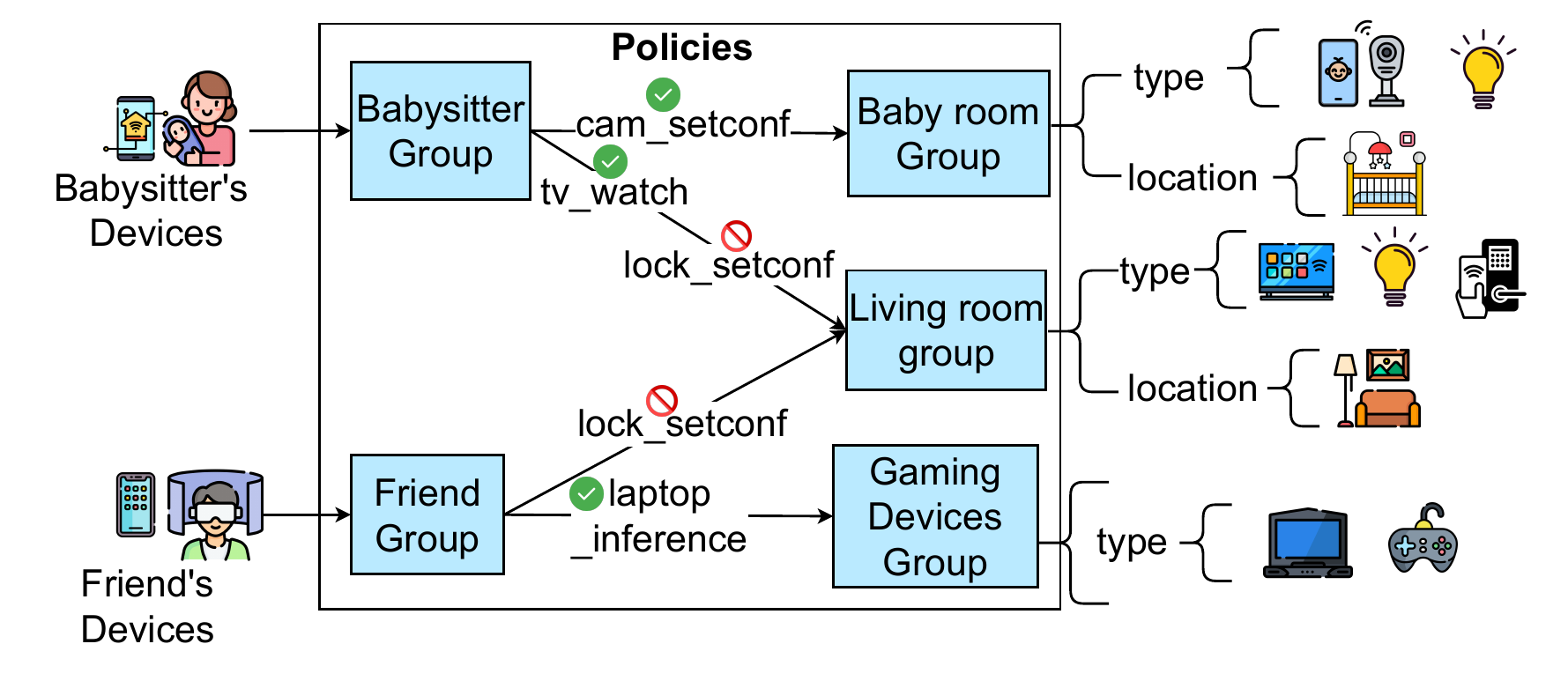}
    \caption{An end-to-end access control policy description, where babysitters are given access to cameras, lights in the baby room, and entertainment devices in the living room, while friends are given access to the gaming console and the living room devices.}
    \label{fig:end_to_end_sub_obj}
\end{figure}

~\autoref{fig:end_to_end_sub_obj} shows an example access control policy with two users, a babysitter, and a gaming friend. We list the policy using our grouping and access control policy scheme in \autoref{fig:policy_spec_example} and discuss in \autoref{sec:implementation}.  
As shown, the grouping policies use location and device type attributes to generate grouping schemes (e.g., devices in the living room). 
The policy engine uses these groups to dynamically select devices matching the attributes at runtime, which implies handling dynamics where guests and native devices may arrive or leave. In addition, the figure shows how policies connect baby room and living room groups with the babysitter groups, enabling the assignment of fine-grained access to the babysitter.

\begin{figure}[t]
    \centering
    \includegraphics[width=\linewidth]{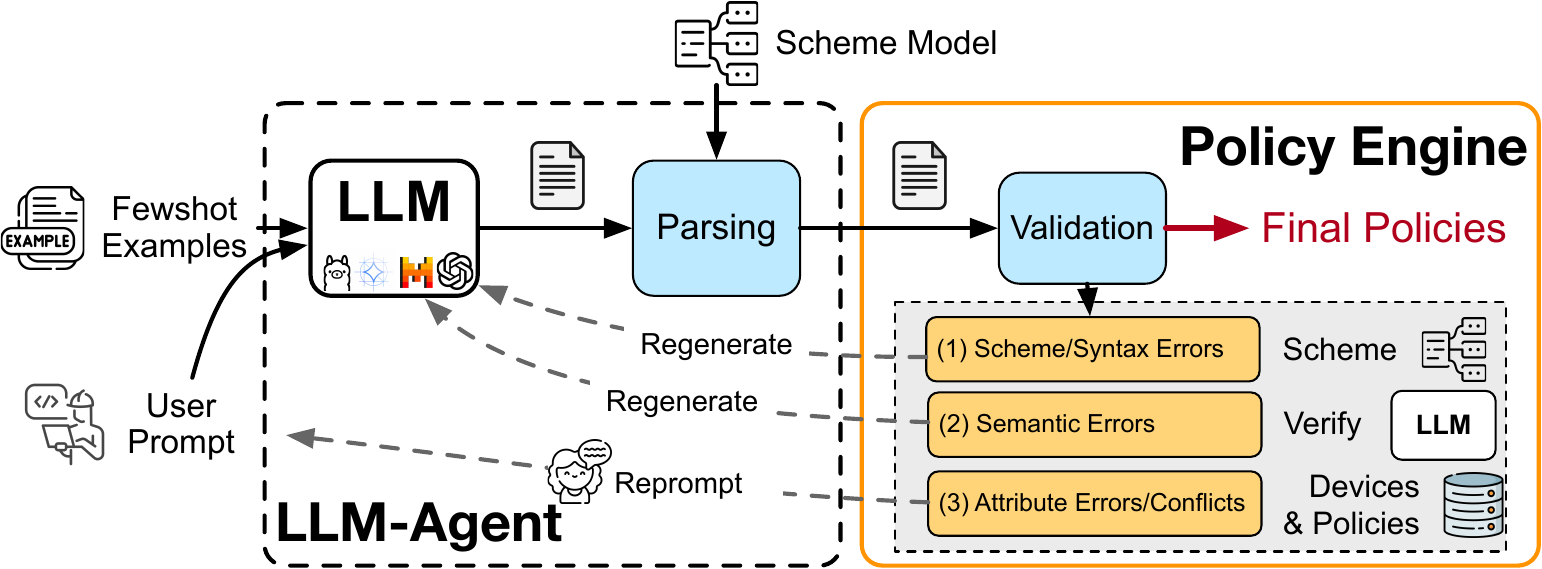}
    \caption{LLM-based policy generation and validation pipeline.}
    \label{fig:llm-pipeline}
\end{figure}

\subsection{LLM Policy Generation Pipeline}\label{sec:des_llm}

Generating access control policies requires a level of experience not commonly available for end-users. For instance, our configuration scheme requires knowledge of YAML and a basic understanding of access control systems. To circumvent such requirements, in \systemName, we depend on LLMs' abilities to generate structured outputs, such as in programming and configuration tasks \cite{subramaniam2024intent, vatsa2025synthesizing, cheng2025:SayWhatYouMean}.  
For example, to define a group for the devices in the baby room, the user can prompt: "Create a group with lights and the baby monitor in the baby's room", or "
All \texttt{door-locks} on the \texttt{east-side} of the building." In addition, to create an access control policy that allows his friends to access the gaming devices, the user can prompt: "Allow \texttt{my-friends} to \texttt{use} the \texttt{gaming-device}".
The key challenge, however, in this situation is that LLMs may not produce syntactically correct outputs, may not represent the user's intent accurately, or may not be compatible with the current system configuration. 

\autoref{fig:llm-pipeline} shows our proposed LLM-based policy generation and validation pipeline. Our LLM-based pipeline consists of two components: an LLM agent that processes user prompts and generates structured and type-safe output. In addition, it contains a validator (part of the policy engine) responsible for validating new configurations.
Our pipeline behaves as follows: When a user issues a configuration prompt, \systemName's LLM-agent augments the user prompt with a context (often referred to as few-shot examples  \cite{Zhao2021:FewShot}) that contains the policy definition and a few example policies based on the policy type which enhances the behavior or the LLM model. 
The LLM agent also instructs the LLM model to produce output that follows a structured and type-safe model scheme.
The structured scheme is defined as a class representation of the YAML file that constrains the format and types of the policies as per policy type (see \autoref{lst:pydanticsheme}). The class representation also includes rules on the allowed attributes and constraints on domain values, as well as other complex validation constraints. 

To validate a new policy, the policy engine employs a three-step validation approach.
First, the policy engine validates the scheme of the generated \texttt{YAML} files against the \systemName schemes. Although the LLM-agent instructs the LLM model to follow the model scheme, we found that since our scheme was not in the training set, LLM may generate a configuration with a scheme error. In this case, \systemName uses a feedback loop with LLM while referring to the mistake that the LLM created based on the custom validators we implemented in the model. 
Second, since LLMs may produce outputs that deviate from users' intentions (i.e., hallucinations), we implement an LLM-based \emph{actor-critic} feedback loop that verifies semantic errors to ensure the original prompt retains the same semantic meaning as the generated prompts. To do so, we convert the generated policy into a text representation using a predefined template and feed it, along with the original prompt, to the LLM, asking it to validate that both hold the same semantics. Similar to the scheme errors, if it fails, we use a feedback loop to address the foreseen issue. Third, the policy engine validates the output against attribute errors (e.g., a user referred to a location not defined in the system) or new policies that bring configuration conflicts. In this case, we send the prompt to the user for correction. Note that in some situations, our agent may not be able to generate valid configurations, for example, when user prompts are unclear. However, in this case, our pipeline can flag their correctness and ask the user for correctness. 
Lastly, once the LLM-generated output is validated, we convert the generated classes into YAML files and feed them to the policy engine database.

\begin{figure*}[t]
    \centering
    \includegraphics[width=0.8\linewidth]{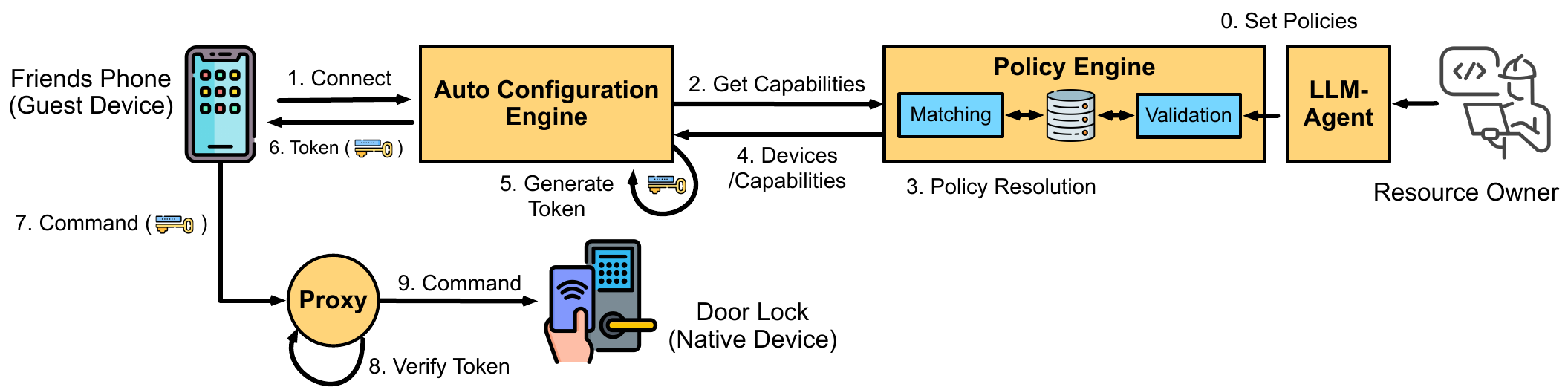}
    \caption{End-to-end workflow, where a friend's phone is given access to the door lock.}
    \label{fig:access_flow}
\end{figure*} 

\subsection{Device Auto Configuration}\label{sec:auto_config} 
When a new device joins the collaboration network, the \systemName automatically configures its permissions based on its attributes and the current policies. We assume that authenticated devices may hold a set of global attributes as part of their authentication certificates (e.g., a delivery robot may authenticate itself via a certificate that indicates and verifies its attributes). Additionally, the system administrator may tag this new device with a set of attributes that are then used for configuring the device (e.g., tagging the device as `friends'), which can also be automated using a discovery LLM-based agent. 

While auto configuration may include multiple operations, we focus on the process of generating access control capability tokens and assigning them to new devices.  
To configure a new device with the proper access control permissions, we rely on cryptographically signed capability-based tokens. We generate the tokens as follows.
After the device joins the network, the auto configuration engine first interacts with the policy engine to find all policies that match the device's attributes, which determines the device's permissions and rate limits accessible to a guest device based on the specified policies.
Each matching policy enumerates all groups mentioned within its policies and determines all matching devices. The auto configuration engine then generates a capability token (also known as an access token) for each matching device, along with the allowed capabilities (i.e., actions that are permitted on the devices), rate limits (e.g., requests per second), and the token's expiration time. The utilized structure of the tokens is presented in ~\autoref{sec:impl_scheme}.
Capability tokens are signed and encrypted by the auto configuration engine's public key. Tokens are then sent to guest devices to enable access to the native devices.

It is important to note that the policy engine generates the capabilities by merging all policies that match this device, which is achieved by combining the policies and the specified devices into a comprehensive list. 
For instance, if the friends' group has two policies allowing them to access all light bulbs and all devices in the living room, they will retain access to all the light bulbs in the living room and other rooms. 
In addition, when the policy changes, the policy engine iterates over the affected groups and their associated native devices, issuing an invalid token request. Device proxies retain a list of invalidated tokens until they expire.

\subsection{Proxy-based Device Collaboration} \label{sec:des_proxy}

In \systemName, devices communicate via lightweight proxies that enable vendor-independent communication, enforce access control, impose rate limits, and augment the capability of accessed devices.  In \systemName each native device in the system has a proxy\footnote{Note that guest devices that provide services (e.g., a friend who brings her gaming console) require a proxy to enforce access control and rate limits.}; to interact with a device, a guest device sends an RPC request to the device's proxy along with the capability token. The proxy first verifies that the token is not in the invalid lists.
Then, the proxy decrypts the request using the public key of the auto configuration engine and verifies the token's validity, capabilities, and rate limits before allowing the request. 
A request is denied if the capability does not permit the requested action or if rate limits are exceeded. For efficiency reasons, signed tokens are validated during session establishment time and cached. The proxy maintains a list of verified tokens in its cache to reduce per-request verification overhead in the data path. Details of token validation are presented in~\autoref{sec:impl_data}.

Each proxy exposes a virtual, hardware-independent interface for its device, enabling device-to-device communication without requiring knowledge of vendor-specific interface details, as these interfaces vary from one vendor to another, even for the same device type. 
For example, all IoT cameras can be accessed using the same abstract interface regardless of the vendor.
For each allowed request, the proxy performs interface translation to convert the request to the underlying vendor-specific interface and forwards the request for execution on the device. The hardware-independent interface in the proxy provides a virtual interface between access requests and specific device APIs. 
~\autoref{tab:cap_att_examples} shows examples of the generic APIs and capability definitions for different device types. 

\subsection{Collaboration Workflow}\label{sec:des_example}
~\autoref{fig:access_flow} lists an example and shows how the resource owner and a guest device interact with \systemName. We assume all point-to-point interactions are encrypted to ensure secure interactions and token distributions.
First, the resource owner uses the LLM to create a one-time dynamic grouping and access control polices which are verified by the policy engine (Step 0). 
When a new guest device is connected and tagged with appropriate attributes when it joins the system for the first time (e.g., a friend's phone), the auto configuration engine contacts the policy engine, which computes the accessible devices and capabilities (Steps 2-4). Subsequently, the auto configuration engine creates a token per native device (Step 5). Then, the auto configuration engine sends this token to the guest device (Step 6) along with the device proxy access information (e.g., IP address and device type). 
Finally, when the guest device issues a command, the native device proxy verifies the token, ensures that the requested command is authorized, and sends it to the device. Note that the interaction between native and guest devices can be \textit{bidirectional}. To enable that, the guest can rely on a \systemName-style control plane (e.g., hosted in the cloud), which can issue capability tokens to other connected devices.

\section{Implementation}
\label{sec:implementation}
This section outlines an implementation prototype of \systemName described in ~\autoref{sec:design}. 
Our \systemName's prototype is implemented in Python3 using $\sim$2.6k SLOC. 
\systemName components' communicate with external devices (e.g., Guest Devices) via GRPC, and unless otherwise stated, all communications are encrypted using SSL/TLS with AES-256.
In the remainder of this section, we outline the implementation of our access control scheme, detail our LLM-based policy generation and validation pipeline, explain our auto-configuration implementation, and finally describe our proxy-based data plane.

\subsection{Access Control Scheme}\label{sec:impl_scheme}
Our schemes comprise a separate dynamic grouping scheme and access control policies. We based our policy specification on NIST Next Generation Access Control (NGAC) architecture \cite{Ferraiolo2011:ThePolicyMachine} and implemented an access control spec using \texttt{yaml}.
Our access control policy implementation allows users to define device groups $G$ based on the logical relationships between devices and their attributes, and access control policies $P$ that connect these groups using the allowed capabilities and rate limits.
Our dynamic grouping scheme is an attribute-based access control scheme that uses hardware-independent attributes to construct groups of accessible devices (see ~\autoref{tab:cap_att_examples}). The grouping scheme defines \texttt{include} and \texttt{exclude} lists per attribute type.
Our specification assumes that items in the same include/exclude lists are \emph{ORed}, while items across lists (i.e., for different attributes) are \emph{ANDed}.
Our policy engine saves the policies in the SQL database that are queried whenever a new guest device arrives. 

 \begin{figure}[t]
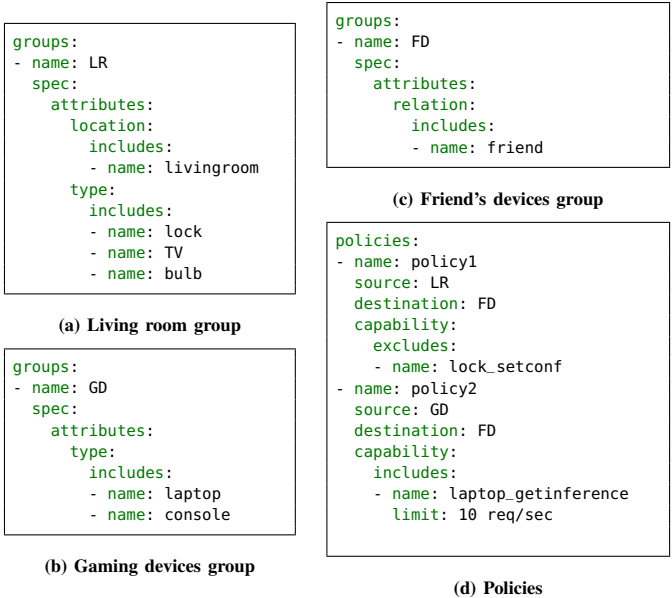

\noindent
\begin{minipage}{0.214\textwidth}
\begin{minted}[frame=single, fontsize=\scriptsize]{yaml}
groups:
- name: LR
  spec:
    attributes:
      location:
        includes:
        - name: livingroom
      type:
        includes:
        - name: lock
        - name: TV
        - name: bulb
\end{minted}
\centering
{\scriptsize \textbf{(a) Living room group}} \\[5pt] %

\begin{minted}[frame=single, fontsize=\scriptsize]{yaml}
groups:
- name: GD
  spec:
    attributes:
      type:
        includes:
        - name: laptop
        - name: console
\end{minted}
\centering
{\scriptsize \textbf{(b) Gaming devices group}} \\[5pt] %
\end{minipage}
\hfill
\begin{minipage}{0.252\textwidth}
\begin{minted}[frame=single, fontsize=\scriptsize]{yaml}
groups:
- name: FD
  spec:
    attributes:
      relation:
        includes:
        - name: friend
\end{minted}
\centering
{\scriptsize \textbf{(c) Friend's devices group}} \\[5pt] %
\begin{minted}[frame=single, fontsize=\scriptsize]{yaml}
policies:
- name: policy1
  source: LR
  destination: FD
  capability:
    excludes:
    - name: lock_setconf
- name: policy2
  source: GD
  destination: FD
  capability: 
    includes:
    - name: laptop_getinference
      limit: 10 req/sec
                        
\end{minted}
\centering
{\scriptsize \textbf{(d) Policies}} \\[5pt] %
\end{minipage}
\caption{Grouping scheme and access control policy to describe the policies illustrated in ~\autoref{fig:end_to_end_sub_obj}.}
\label{fig:policy_spec_example}    
\end{figure}

~\autoref{fig:policy_spec_example}a lists the living room devices (see~\autoref{fig:end_to_end_sub_obj}) where the resource owner creates a group that selects ``locks, TVs, and bulbs in the living room''. ~\autoref{fig:policy_spec_example}a lists the living room devices (see~\autoref{fig:end_to_end_sub_obj}) where the resource owner creates a group that selects ``locks, TVs, and bulbs in the living room''. ~\autoref{fig:policy_spec_example}b and ~\autoref{fig:policy_spec_example}c define the groups for gaming devices and the friends group. 
~\autoref{fig:policy_spec_example}d lists two access control policies from the friends' scenario in ~\autoref{fig:end_to_end_sub_obj}. \texttt{policy1} permits friends to access the living room devices and all capabilities except for setting the lock configuration. In contrast, \texttt{policy2} permits the friends to access the inference service with a maximum rate of 10 req/sec.

\subsection{\systemName LLM-based Policy and Validation Pipeline}
\begin{lstlisting}[float=tp,style=mypython, caption={Pydantic schema for group schema model.},label=lst:pydanticsheme]
class NameEntry(BaseModel):
    name: str

AttributeTypeLiteral = Literal[
    "deviceLocation", "deviceType", "deviceStatus",
    "deviceManufacturer", "ownerName", "deviceBattery"
]

class AttributeSpec(BaseModel):
    includes: Optional[List[NameEntry]] = None
    excludes: Optional[List[NameEntry]] = None

class Spec(BaseModel):
    attributes: Dict[AttributeTypeLiteral, AttributeSpec]

class GroupItem(BaseModel):
    groupName: str
    spec: Spec

class GroupGrammar(BaseModel):
    groups: List[GroupItem]
\end{lstlisting}

Our policy generation pipeline uses a Large Language Model (LLM) to compile user intents into a grouping scheme or access control policies. 
Our pipeline comprises an LLM agent responsible for generating a structure and type-safe configuration. We build our agent on top of the Pydantic LLM agent framework \cite{pydanticai} and use models supplied by Ollama\cite{ollama}. Pydantic AI utilizes Pydantic validation models, which provide an intuitive way of self-validating schemes. We formalize both group and access control policies using Pydantic validation models. \autoref{lst:pydanticsheme} shows an example of a schema that validates a grouping scheme and lists optional properties and domain values for attributes. 

At runtime, our LLM agent takes the user prompt, the validation model, and a few-shot prompting examples (denoted as 
\texttt{system\_prompt}) and produces results that follow the examples and the validation model. These validation models are passed to the LLM to constrain the output and ensure type-safe and structured generation. The LLM output is automatically parsed and validated against this model, enabling us to identify and correct type-missing fields or malformed outputs early. Moreover, we implement a feedback loop to reprompt the LLM in case of syntax or scheme errors as explained in \autoref{sec:des_llm}.
Additionally, we implement a semantic validation agent that ensures the generated configuration aligns with the user's intent. The validation process works by converting the generated configuration into a text representation and comparing it with the original prompt, asking the agent to confirm that they hold the same semantics. Lastly, our attribute validation is implemented using Python, which queries the policy and devices database to validate attributes and resolve conflicts.

\subsection{\systemName Automatic Configuration}\label{sec:impl_conf}
\systemName operates as a transparent gateway (i.e., home hub) and is able to detect new connections, issuing our auto-configuration protocol that spans the policy and auto configuration engine and provides the new device with an access token and a list of accessible devices. When a new device is authenticated, the auto configuration engine then interacts with the policy engine and computes the accessible devices, the allowed capabilities, and the rate limits for each device, as well as generates an access token for each device. Our system encodes these capabilities as JSON Web Tokens (JWT)~\cite{JWTSpec}. Our tokens use official attributes (also called registered claims) and custom (private) claims such as \texttt{inc\_cap}, which lists the allowed capabilities and resource limits as described in~\cite{9833873}.

The following example explains different claims and shows how we provide \texttt{VR\_headset} with capabilities to offload processing requests but not to update the gaming console configurations. 
\begin{lstlisting}[language=json, label=lst:token]
{
 "iss": 'CollabIoT', //Issuer (CollabIoT)
 "sub": 'VR_headset', //Subject(Guest Device)
 "aud": 'Gaming_laptop', //Audience(Native Device)
 "iat": 1731425174, //Issued At Time
 "nbf": 1731425174, //Not Before Time
 "exp": 1731459572, //Expiration Time
 "inc_cap": // Included Capabilities and Limits
    [{"offload":50}], 
}


\end{lstlisting}

After creating the tokens, the auto configuration engine encrypts them using ED25519 256 and signs them using EdDSA digital signatures.
These tokens are validated before executing the requests or starting a communication session. Our implementation assumes that the client is authenticated by validating its certificate using a certificate authority, bearer tokens, or group signatures~\cite{boneh2004:shortgroup}, which is practical in scenarios such as the delivery robot example, where companies own a fleet of devices.

\subsection{Proxy-based Device Collaboration}\label{sec:impl_data}

The device proxy implements a hardware-independent GRPC interface for its device type (see ~\autoref{tab:cap_att_examples}). It encapsulates the hardware-independent interfaces by translating the guest device's request to the hardware-dependent interface or utilizing a device-specific SDK (e.g., Wyze SDK~\cite{wyze_sdk}). The proxy uses asynchronous communication to serve concurrent access requests. We implement the rate limit with a token bucket rate limiter~\cite{tang1999network}. 
For ease of deployment, we implement device proxies as Docker containers, assigning one CPU core per proxy instance.

\section{Evaluation}
\label{sec:evaluation}
In this section, we evaluate the efficacy of \systemName using an IoT testbed and various case studies. We start by describing our experimental setup and then present our case studies that shows the end-to-end behavior of \systemName. Then we evaluate various components of \systemName.  
Lastly, we demonstrate the performance of the \systemName control plane and data plane with multiple micro-benchmarks and large-scale emulations. 

\begin{figure}[t]
    \centering  
    \includegraphics[width=0.9\linewidth]{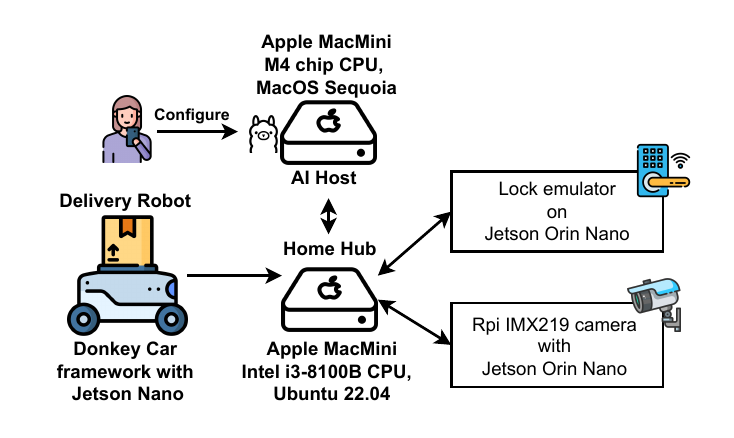}
    \caption{Experimental testbed.}
    \label{fig:robot_exp}
\end{figure}

\begin{figure}[t]
    \centering  
    \includegraphics[width=0.9\linewidth]{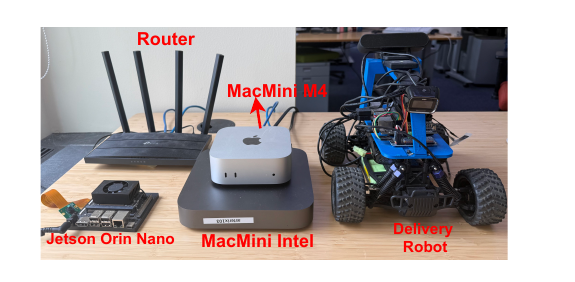}
    \caption{\systemName real testbed.}
    \label{fig:iot_testbed}
\end{figure}

\subsection{Experimental Setup}\label{sec:eval_setup} 
In this section, we detail our experimental setup, LLM models, and IoT devices.

\noindent\textbf{Hardware Testbed.}
Our testbed, shown in \autoref{fig:robot_exp} and \autoref{fig:iot_testbed}, is based on a local IoT testbed. The testbed uses an Apple Mac mini as a home manager hub. The Mac mini has an Intel i3-8100B CPU running Ubuntu 22.04. The home manager hub hosts \systemName's control plane and the data plane. 
In addition, we utilize a Mac mini with an M4 CPU to host local LLM models from ollama~\cite{ollama} and our LLM agent and emulate IoT devices in our large-scale experiments.

\noindent\textbf{LLM Agent.}
Our LLM agent can use multiple local and cloud LLM models (See \autoref{tab:llm_policy_generation_time}). We evaluate the behavior of different models under two scenarios, each with 20 grouping and 20 access control policies. Our first scenario, which we denote as \texttt{structured prompts}, assumes that the user follows a very elaborate format, which details the grouping or access control policy. Our second set of prompts, which we denote as \texttt{unstructured prompts}, represents situations where users provide unclear or concise prompts, which we acquire by manually rephrasing the 40 policies.  

\noindent\textbf{IoT Devices.} 
Our testbed, shown in \autoref{fig:robot_exp} and \autoref{fig:iot_testbed}, includes a set of IoT devices:  Raspberry Pi IMX219 camera,
a delivery robot, and a Jetson Orion Nano.
The Delivery Robot is based on the DonkeyCar framework~\cite{donkeycar} and comprises a Jeton Nano developer kit. The Jetson Orion Nano is used to host the AI service (developed using TensorRT)\footnote{https://github.com/dusty-nv/jetson-inference} and emulate different IoT devices (e.g., a door lock). 
Lastly, we also developed sensor and device simulation clients that we deployed on the M4 Mac mini for system scalability evaluations.

\subsection{End-to-End Case Studies}
We implemented two case studies to demonstrate the effectiveness and performance of \systemName.

\begin{figure}[t]
    \centering
    \includegraphics[width=0.9\linewidth]{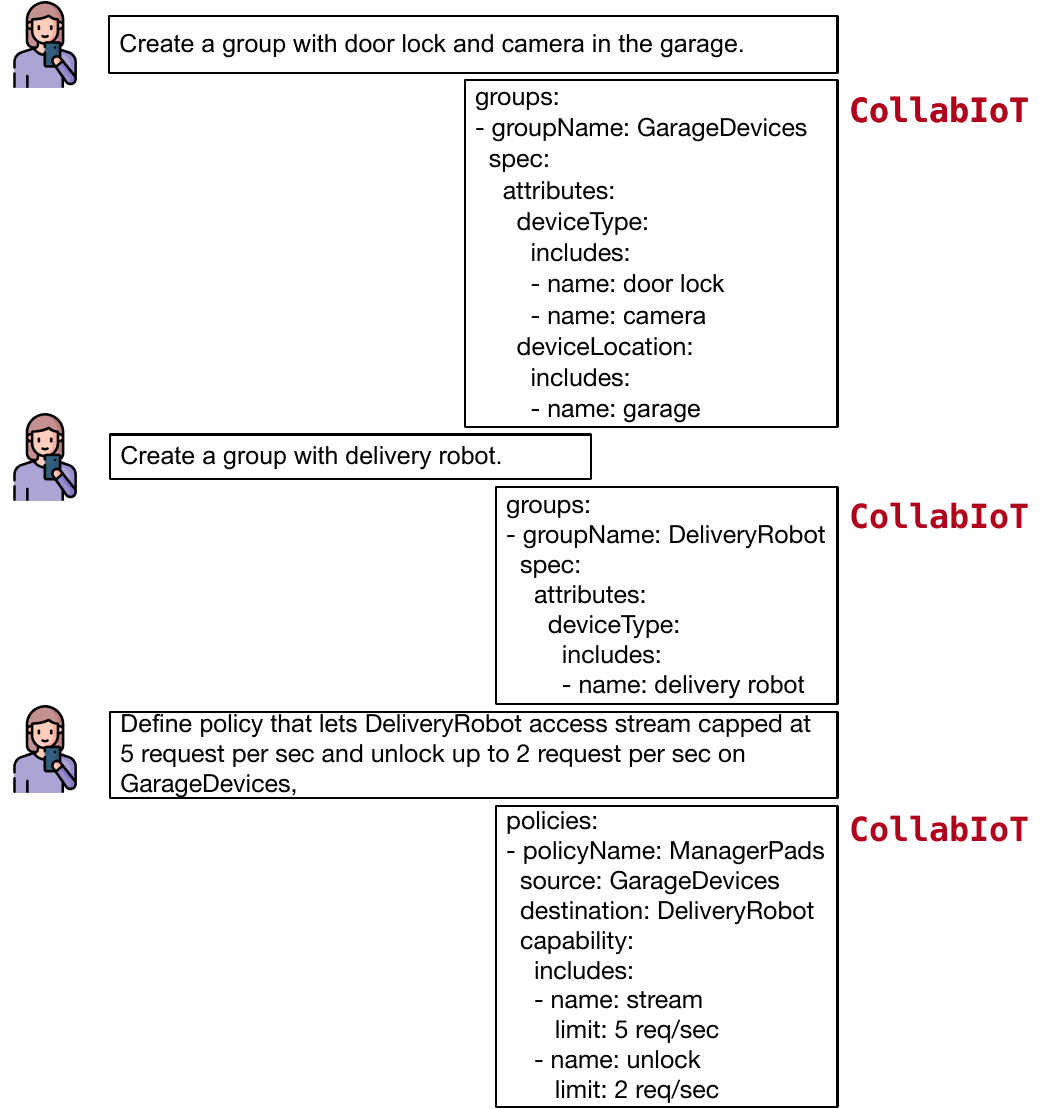}
    \caption{Delivery robot policy configuration.}
    \label{fig:robot_demo}
\end{figure}

\begin{table}[t]
\caption{Results of Case Study 1 (Delivery Robot). }
\label{tab:delivery_robot}
\resizebox{0.9\linewidth}{!}{%
\begin{tabular}{lcc}
    \toprule
    \textbf{Event} & \textbf{Time (ms)} & \textbf{Verification Time (ms)} \\ \toprule
    \textbf{Robot Entering} & 151.51 & NA \\
    \textbf{Lock Access} & 43.22 & 1.07 \\
    \textbf{Camera Access} & 85.36 & 1.13 \\
    \bottomrule
\end{tabular}
}
\end{table}

\subsubsection{\textbf{Case Study 1 - Delivery Robot}}
Our first case study demonstrates our auto configuration and device collaboration abilities by enabling a delivery robot to access the garage door lock and camera. 
In this experiment, the delivery robot (i.e., our donkey car) arrives at a house with a package to be delivered to a designated corner. The robot then requests access to the garage door lock and camera, which aids it in its path-planning decisions.\footnote{We do not implement a path-planning approach but demonstrate the feasibility of accessing the camera live stream.}
\autoref{fig:robot_demo} demonstrates how a user can prompt and generate the access control policies for the delivery robot. In this case, the user issues a prompt to create two groups, one for garage devices and the other for delivery robots. Then, prompt the system to create an access control policy that links both groups, allowing the robot to stream the garage camera at a rate of 5 frames per second and access the garage door twice to unlock and lock the door. 

~\autoref{tab:delivery_robot} shows the steps and the time taken by each step in this scenario, which does not require any human intervention --- except for setting the policies, which happens beforehand. 
As shown, the results demonstrate the responsiveness of \systemName, where it only takes 151.51 ms to connect to the control plane, evaluate the access control policies, list the devices, generate capabilities, and encrypt the access control tokens (Steps 1 to 6 in ~\autoref{fig:access_flow}). In addition, the results depict the end-to-end responsiveness of the data plane, where it takes 43.22 ms to execute the unlock door command and 85.36 ms to access the camera stream. In contrast, in both cases, it takes $\sim$1 ms to verify the access token. Lastly, we note that our policy generation time is model and deployment dependent, \autoref{tab:llm_policy_generation_time} presents the average policy generation time across all promotes, which we detail later in this section.

\begin{figure*}[t]
    \centering
    \begin{subfigure}[t]{0.3\textwidth}
    \centering
        \includegraphics[width=\linewidth]{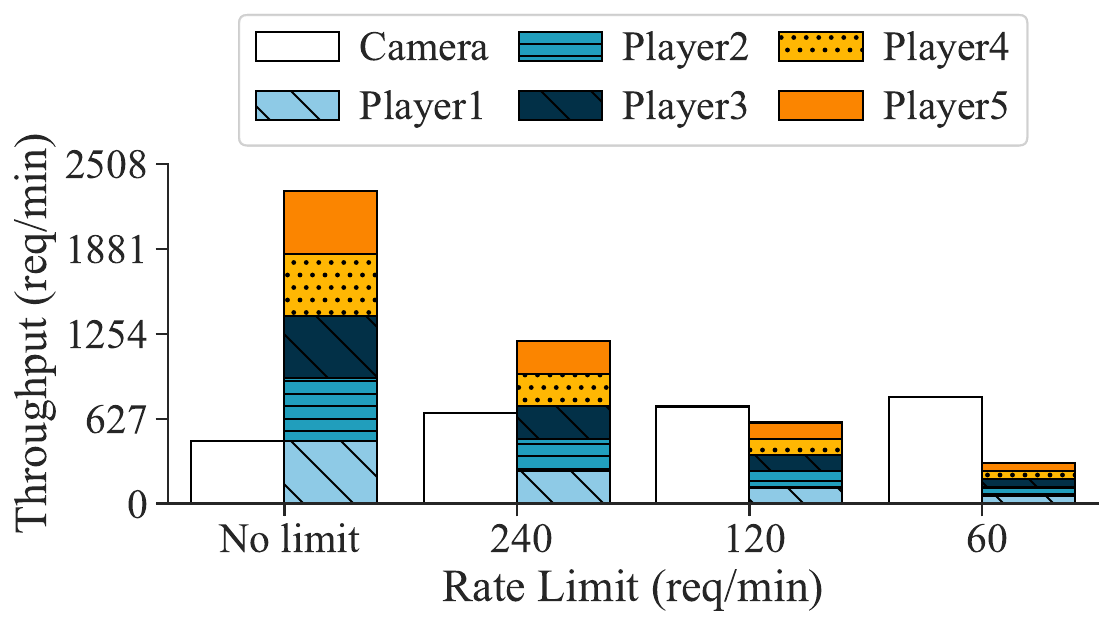}
        \caption{Fixed Rate}
        \label{fig:eval_cs2_limit_static}
    \end{subfigure}
    \begin{subfigure}[t]{0.3\textwidth}
    \centering
    \includegraphics[width=\linewidth]{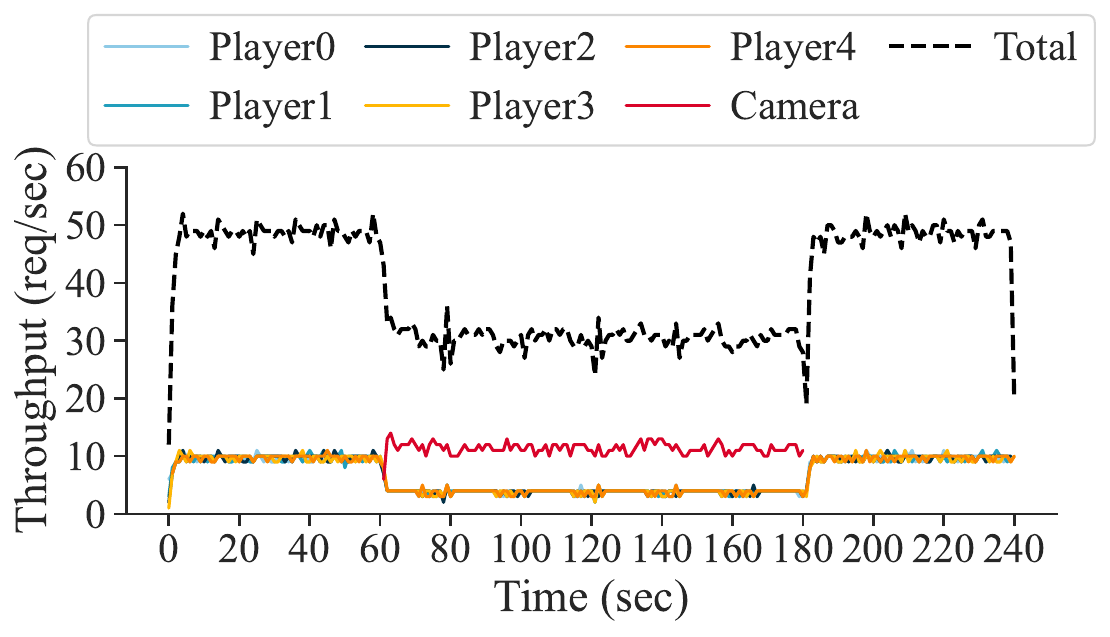}
    \caption{Context-aware rate limits.}
    \label{fig:eval_cs2_limit_context}
    \end{subfigure}
        \begin{subfigure}[t]{0.3\textwidth}
    \centering
    \includegraphics[width=\linewidth]{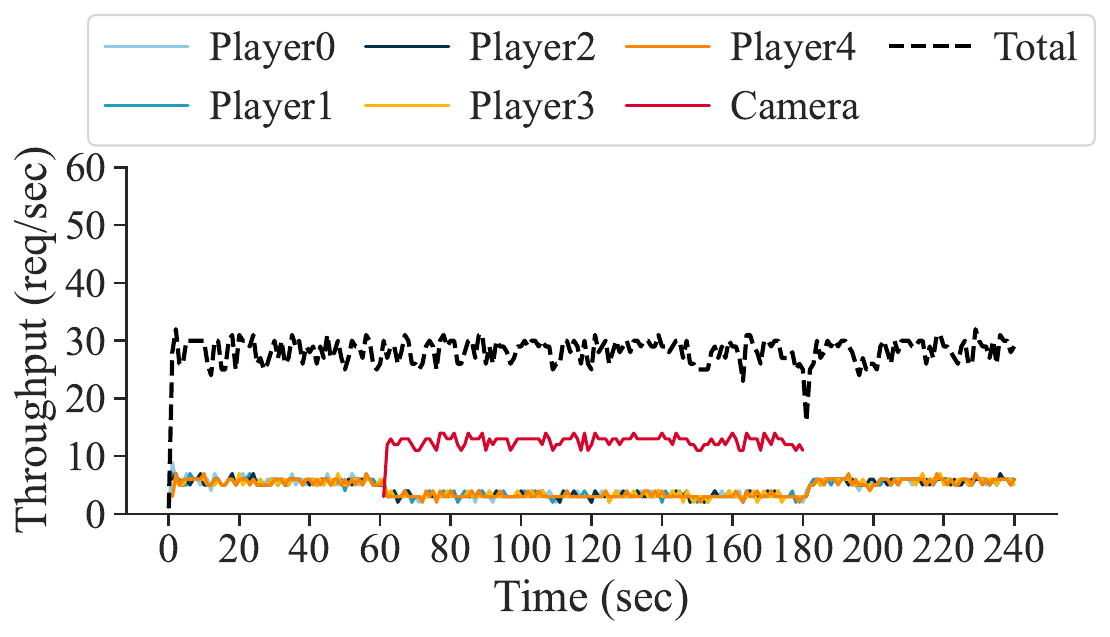}
    \caption{FCFS with priority}
    \label{fig:eval_cs2_limit_FCFS}
    \end{subfigure}
    \caption{Applying rate limits on an AI processing service in different settings. 
    }
    \label{fig:eval_cs2_limit}
\end{figure*}

\begin{figure}[t]
    \centering
    \includegraphics[width=0.6\linewidth]{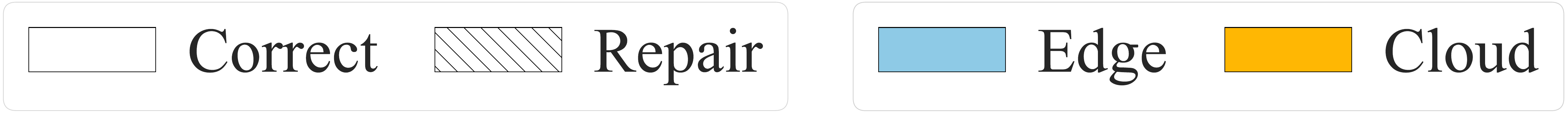}\\
    
    \subfloat[\centering Grouping Policy]{
        \includegraphics[width=0.49\linewidth]{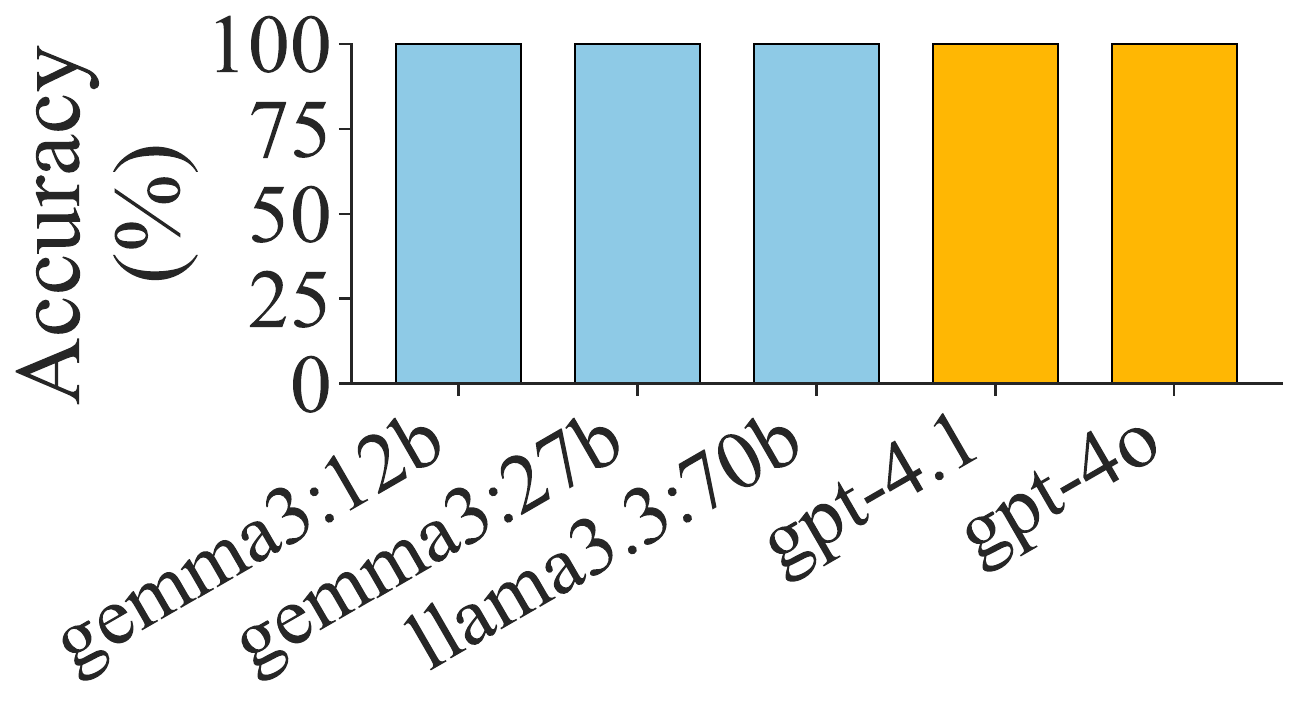}
        \label{fig:llm_struct_group}
    }
    ~
    \subfloat[\centering Access Control Policy]{
        \includegraphics[width=0.49\linewidth]{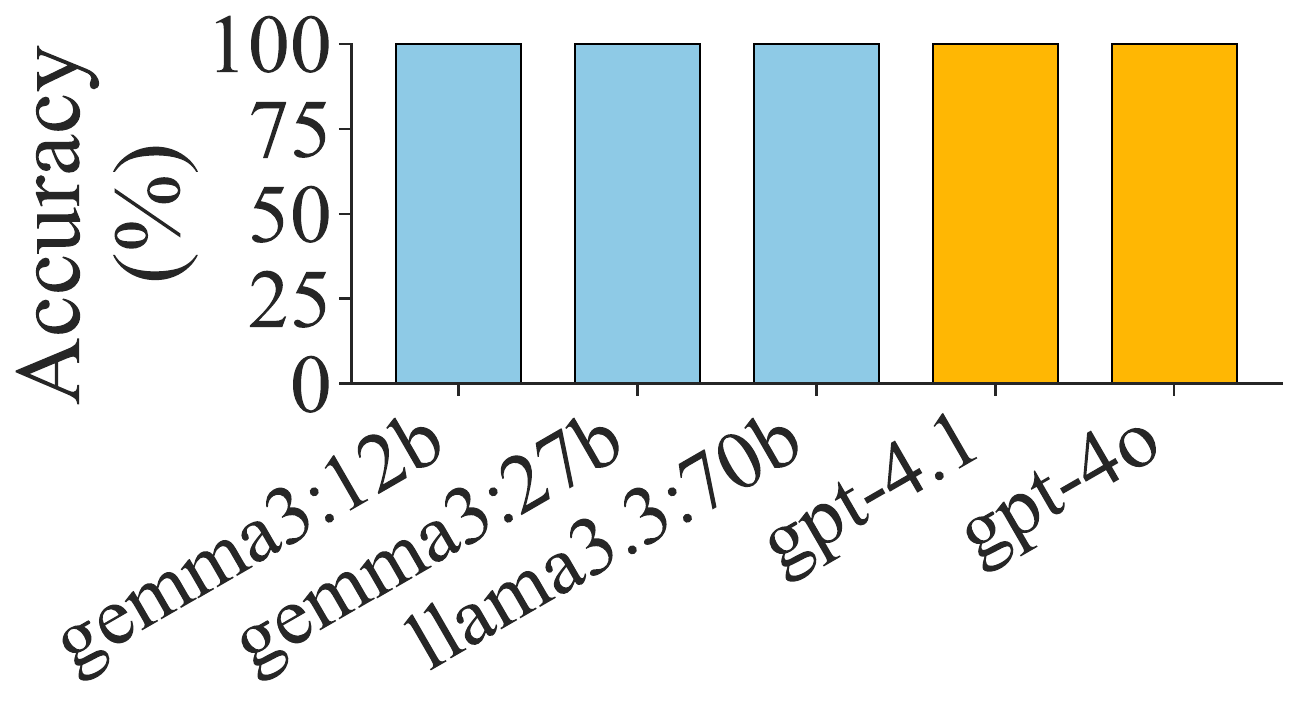}
        \label{fig:llm_struct_policy}
    }
    \caption{LLM agent performance against structured prompts in (a) grouping and (b) access control policies.}
    \label{fig:llm_struct}
\end{figure}

\subsubsection{\textbf{Case Study 2 - AI Processing}}
In our second case study, we focus on \systemName's proxy-based device data plane and demonstrate its ability to augment device capabilities by implementing device-agnostic scheduling and rate control capabilities in the proxy. In this scenario, we consider a GPU-based AI inference service deployed on a Jetson Orin and shared between multiple users in the environment. 
The AI server provides a service utilized by five players (e.g., AR/VR devices engaged in a video game) and a live feed from a security camera, where all services send requests in a closed-loop manner. For simplicity, we assume that all requests are processed using an inference service that runs the ResNet152 DNN model and is deployed with the TensorRT inference framework.  
~\autoref{fig:eval_cs2_limit} shows how \systemName can have implemented device-agnostic rate limiting and scheduling policies:

\noindent\textbf{Static Rate.} \autoref{fig:eval_cs2_limit_static} illustrates the scenario where the proxy enforces a static rate-limiting policy on the players, but imposes no limits on the camera feed. As shown, removing the rate limit allows all players and the camera to push the same number of requests --- total throughput is 2,769, while each player/camera sends $\sim$461 requests/min. However, as we apply more constrained limits, the camera feed becomes more favored, increasing its throughput by up to 790 requests per minute, when all players have a rate limit of 60 requests per minute. Note that since the camera cannot substitute for the requests of all the players, the total request rate shrinks as devices have more aggressive rate limits. 

\noindent\textbf{Dynamic Rate.} Our second scheduling policy demonstrates dynamic rate limiting, which can be used to implement situational awareness by adjusting the rate limits (e.g., in response to camera movement detection). In this case, when the AI model detects any objects, the proxy applies a stricter rate limit on other players, allowing the camera to maintain its rate limits.  ~\autoref{fig:eval_cs2_limit_context} shows our evaluation for four minutes, where we inject the movement detection at 60 seconds for 120 seconds. As shown, before the movement detection, each player can send $\sim$10 requests per second, totaling 50 requests per second. However, when the movement detection is injected, the player's rate is dropped to $\sim$4 requests per second, and the camera rate is maintained at $\sim$12 requests per second, totaling $\sim$30 requests per second.
Note that, similar to the previous example, the cameras are unable to send enough requests to maintain the same throughput when players' devices are limited. 
Finally, when the system state is restored, the proxy can restore the earlier request rates. 

\noindent\textbf{Priority Scheduling.}
Our third policy shows \systemName's ability to implement priority scheduling, where the camera maintains a higher priority than the players. In this case, the proxy features a priority queue that uses the First Come, First Served (FCFS) algorithm for tie-breaking. 
~\autoref{fig:eval_cs2_limit_FCFS} illustrates our evaluation scenario, where, similar to the previous example, the camera issues requests after 60 seconds for 120 seconds, during which the players' requests undergo multiple changes. As shown, while the total throughput remains constant, primarily limited by the FCFS behavior. The camera can be prioritized among other players' devices once it starts issuing requests and can maintain the highest request rate (of $\sim$12 requests per second).

\begin{tcolorbox}[enhanced,size=fbox,drop shadow southwest,sharp corners, colback=white]
\textit{\textbf{Key Takeaways.} 
Our case studies show that \systemName enables users to employ granular access control policies without requiring any technical knowledge, showcasing \systemName's ability to facilitate seamless communication between IoT devices, as well as our proxy-based data plane capabilities to augment devices with new features.
}
\end{tcolorbox}
\vspace{-0.2cm}

\begin{table}[t]
\centering
\caption{The models and sizes incorporated in our LLM agent and their average policy generation time.}
\begin{tabular}{@{}lccc@{}}
\toprule
\textbf{Model}          & \textbf{Parameters} & \textbf{Deployment} & \textbf{Time (s)} \\
\midrule
Gemma 3 ~\cite{team2024gemma}                 & 12B                 & Local               & 7.07              \\
Gemma 3 ~\cite{team2024gemma}                 & 27B                 & Local               & 14.83             \\
LLaMA 3.3 ~\cite{grattafiori2024llama}               & 70B                 & Local               & 22.41            \\
GPT-4.1 ~\cite{achiam2023gpt}                &             & Cloud               & 1.22             \\
GPT-4o ~\cite{hurst2024gpt}                 &            & Cloud               & 1.94             \\
\bottomrule
\end{tabular}
\label{tab:llm_policy_generation_time}
\end{table}

\begin{figure}[t]
    \centering
    \includegraphics[width=0.6\linewidth]{figures/Evaluation/LLM/legend.pdf}\\
    
    \subfloat[\centering Grouping Policy]{
        \includegraphics[width=0.49\linewidth]{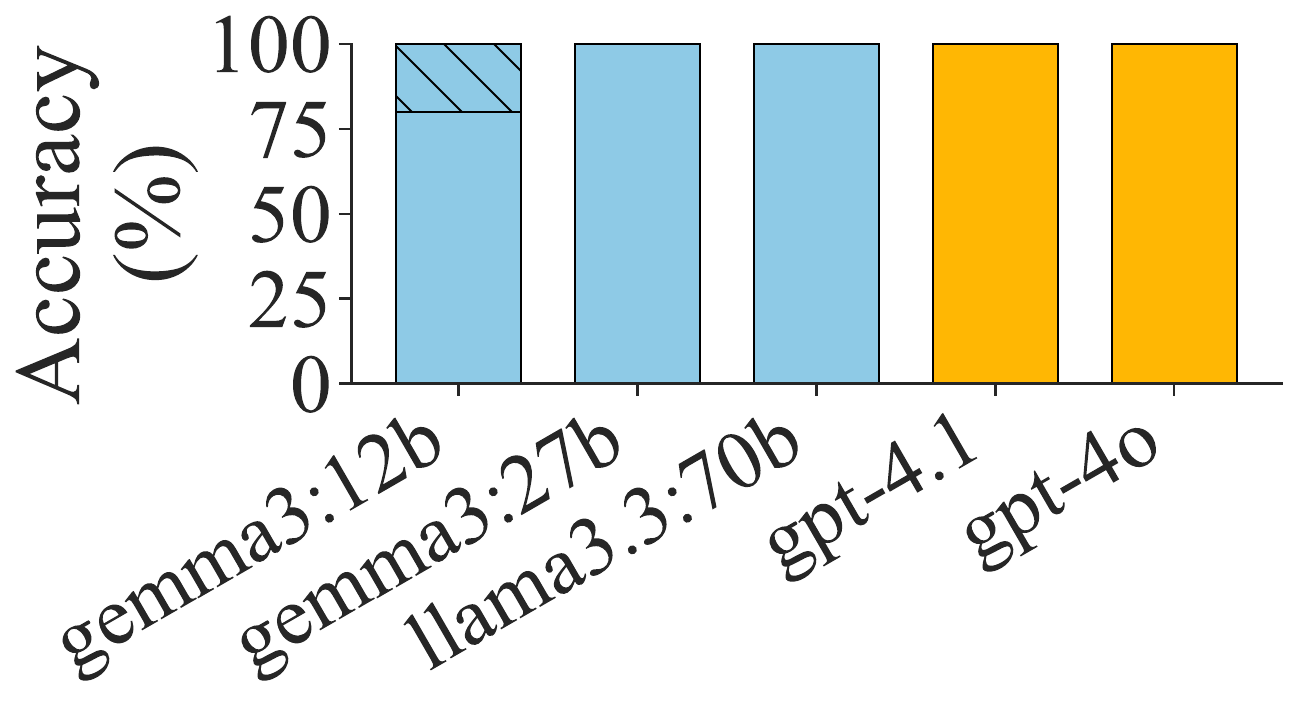}
        \label{fig:llm_unstruct_group}
    }
    ~
    \subfloat[\centering Access Control Policy]{
        \includegraphics[width=0.49\linewidth]{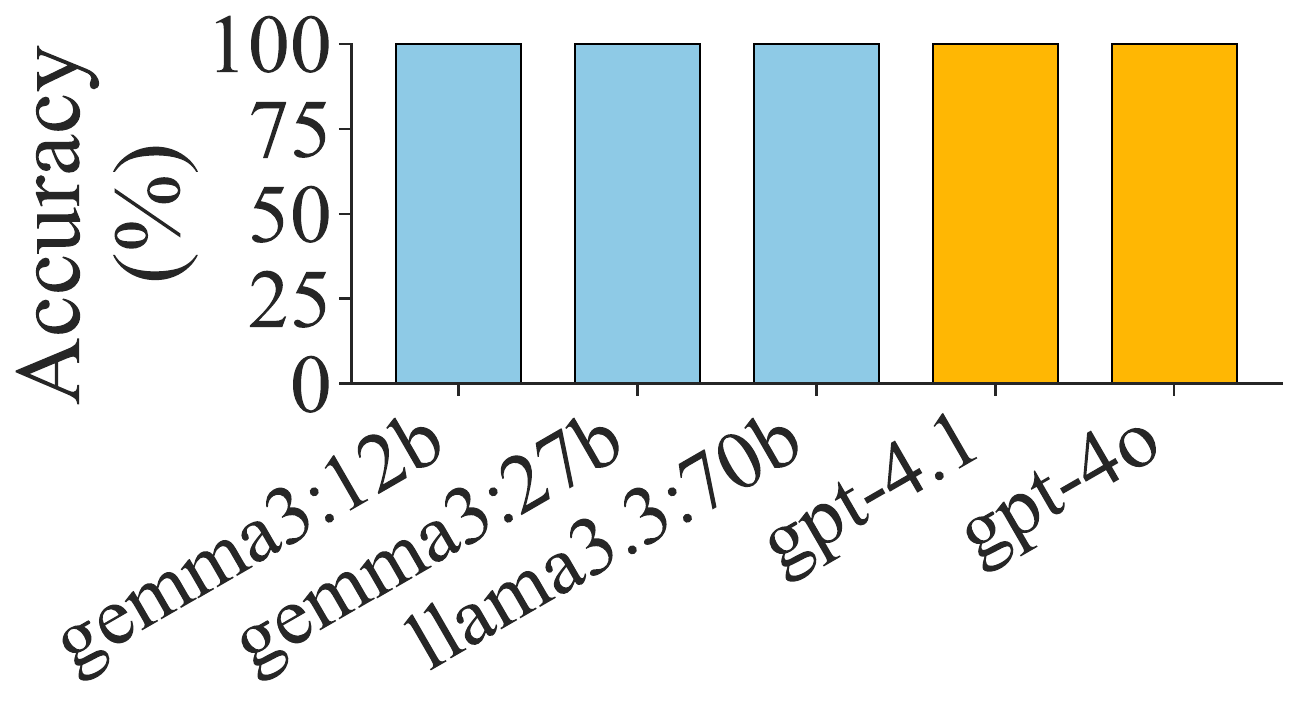}
        \label{fig:llm_unstruct_policy}
    }
    \caption{LLM agent performance against unstructured prompts in (a) grouping and (b) access control policies.}
    \label{fig:llm_unstruct}
\end{figure}

\subsection{LLM-based Policy Generation}

In this section, we address the following question? \emph{Are LLMs able to create valid fine-grained access control policies?} In particular, we evaluate our LLM-based policy generation and validation pipeline. As mentioned in \autoref{sec:eval_setup}, our LLM-agent is based on ollama, which enables us to evaluate multiple models using structured and unstructured prompts, each with 20 grouping
and 20 access control policies.  \autoref{tab:llm_policy_generation_time} presents five LLM models that we utilized in our evaluation, ranging from 12B to 70B open-source and local models, as well as paid cloud models\footnote{GPT-4* is a mixture of experts model, whose size is not publicly disclosed.}. Note that we use each model across the entire pipeline (i.e., the same model is used both in the generation and validation parts, but with different contexts). As our goal was to produce structured output, the table highlights how the model size and deployment can affect the processing time. For instance, a small model, such as Gemma 3 (12B), takes around 7.07 seconds to produce a response, while large models deployed locally can take up to 22.41 seconds. However, the cloud models, although they have a larger size, respond faster because they utilize fast accelerators.

\autoref{fig:llm_struct} illustrates the behavior of the LLM policy generation pipeline, sorted by model size, when the user provides well-structured prompts (e.g., \texttt{Create a group including devices where type is thermostat, location is bedroom and living room.}). As shown, all models are consistently able to generate functional configurations correctly for both grouping and access control policies. 

\autoref{fig:llm_unstruct} evaluates the behavior of our LLM pipeline with scenarios where the user uses a less structured prompt (e.g., \texttt{Create a group with oven, microwave, and fridge but leave out the ones in the garage.}). Similar to \autoref{fig:llm_struct}, the results show that all models generate correct configurations for access control policy. However, for grouping policy the results highlight that Gemma 3 (12B) fails to generate correct result for 20\% times but due to the validation and feedback loop it repairs and gains 100\% accuracy. Intuitively, in comparison to access control policy, generating output from an unstructured prompt for grouping policy is difficult, as LLM needs to classify values into different attribute types. For example, when considering a prompt such as: \texttt{Create a group with plugs in the backyard but leave out the broken ones}, the LLM needs to classify `bulb', `backyard', and `broken' into different attribute types on its own.

\begin{tcolorbox}[enhanced,size=fbox,drop shadow southwest,sharp corners, colback=white]
\textit{\textbf{Key Takeaways.} 
Our results show our policy generation pipeline boosts the performance of edge-based LLMs, making them comparable to cloud-based LLMs. In either case, our results demonstrate the feasibility of generating accurate and functional grouping and access control policies. Our results highlight the value of the validation feedback pipelines in detecting and correcting issues in the generated configurations, especially for unstructured prompts.
}
\end{tcolorbox}

\subsection{Micro-benchmarks}
In addition to the end-to-end evaluations, we implement micro-benchmarks to evaluate the performance and scalability of our \systemName' control and data planes. To ensure consistent results, we repeat each experiment 20 times and plot the average.

\subsubsection{\textbf{Policy Engine}} 
\begin{figure}[t]
    \centering
    \begin{subfigure}{0.48\linewidth}  
    \centering
    \includegraphics[width=\linewidth]{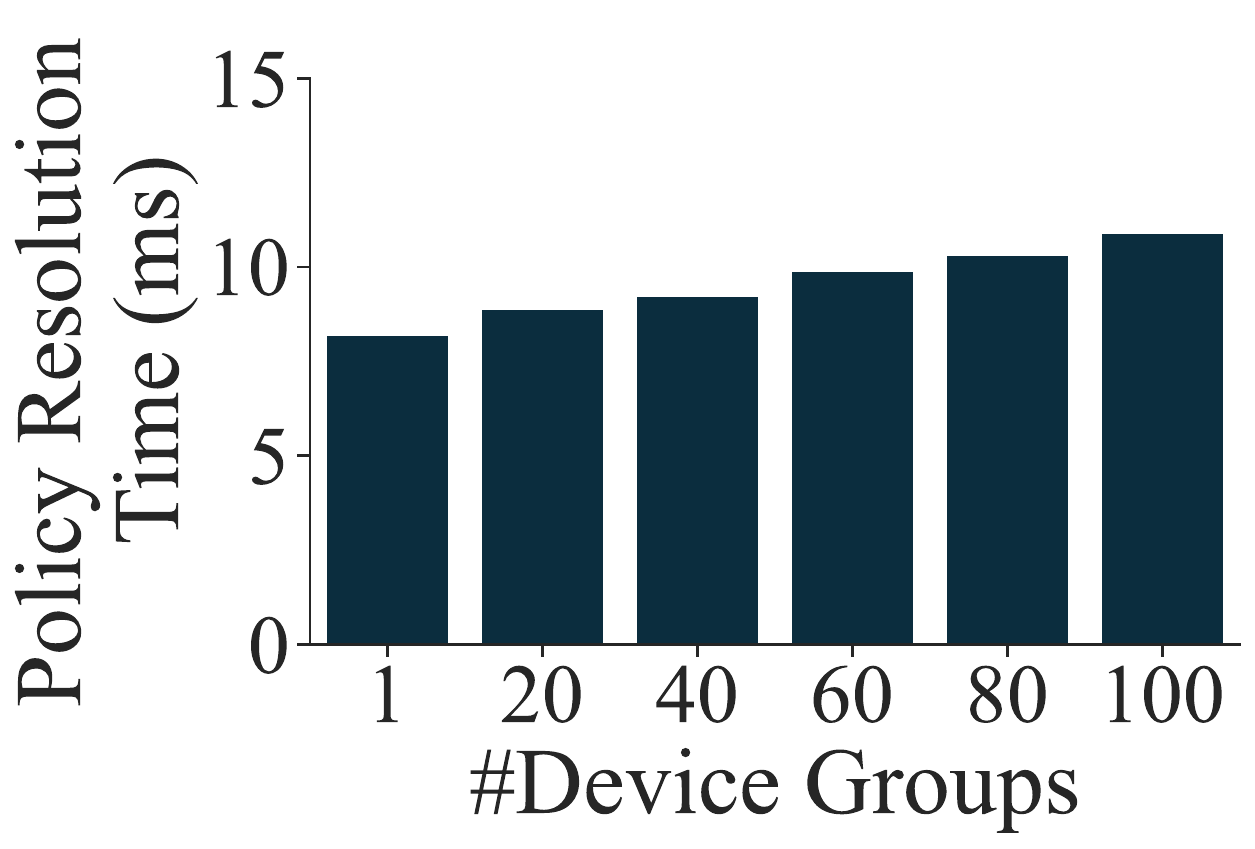}
    \caption{Varying Device Groups}
    \label{fig:policy_engine_res_subj}
    \end{subfigure}
    \hfill
    \begin{subfigure}{0.48\linewidth}
    \centering
    \includegraphics[width=\linewidth]{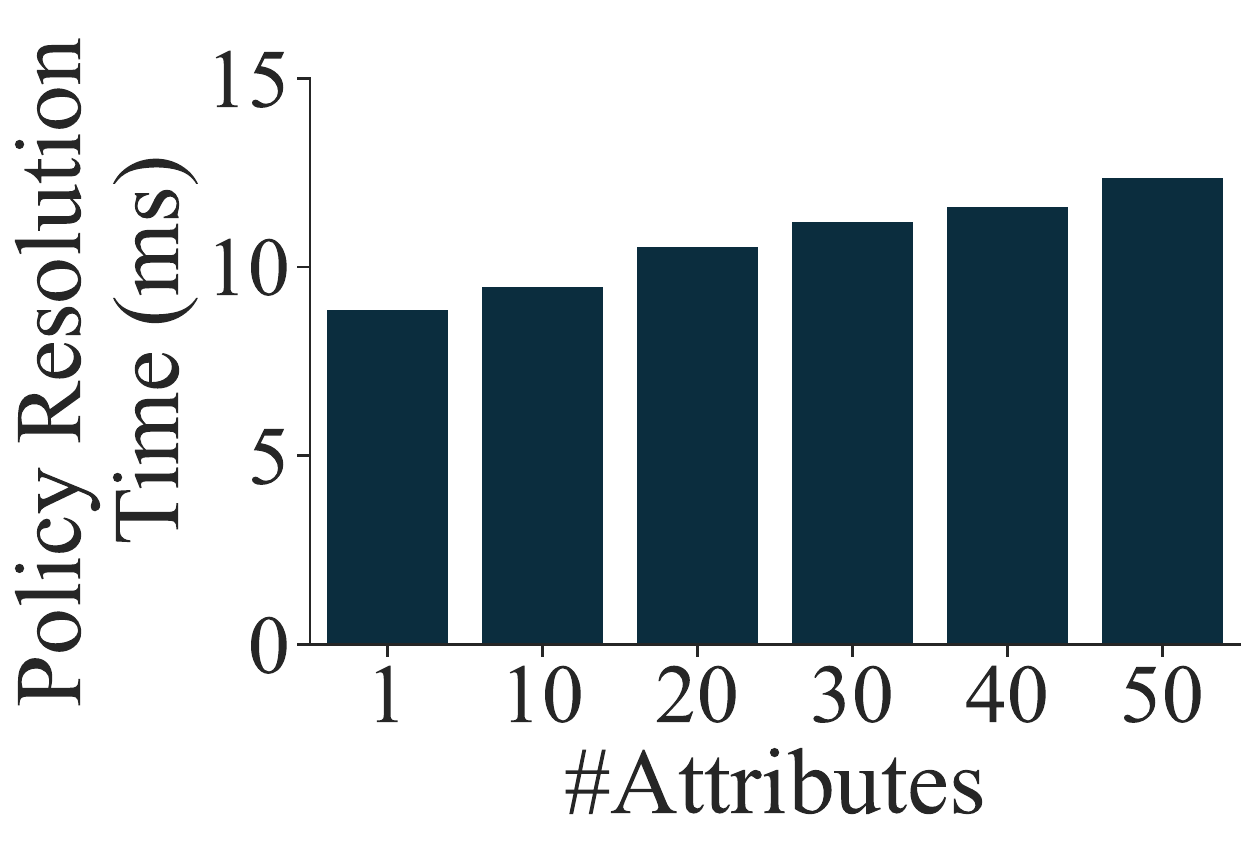}
    \caption{Varying Attributes}
    \label{fig:policy_engine_res_att}
    \end{subfigure}
    \hfill
    \caption{Evaluating policy resolution time when varying the number of groups and attributes.}
    \label{fig:policy_engine_res}
\end{figure}

~\autoref{fig:policy_engine_res} depicts the policy resolution time (Steps 3.1 to 3.4 in ~\autoref{fig:access_flow}) across different numbers of device groups and attributes. 
~\autoref{fig:policy_engine_res_subj} shows the resolution time across different numbers of groups. 
In this case, we create a list of groups, each with ten attribute combinations and a device with 7 attributes , and record the time to compute the groups this device is allowed to access.
As shown, increasing the number of groups has minor effects on the performance, where going from 1 to 20 groups only adds 0.69 ms, while going from 20 to 100 device groups only adds 2 ms. 
~\autoref{fig:policy_engine_res_att} shows how the number of attributes inside a group affects the policy resolution time, where we create a list of 50 groups with combinations of different numbers of attributes from 1 to 50.
Similarly, changing the number of attributes has minor effects, where the difference between a device with 1 and 50 attributes is only 3.52 ms.

\begin{figure}[t]
    \centering
    \includegraphics[width=\linewidth]{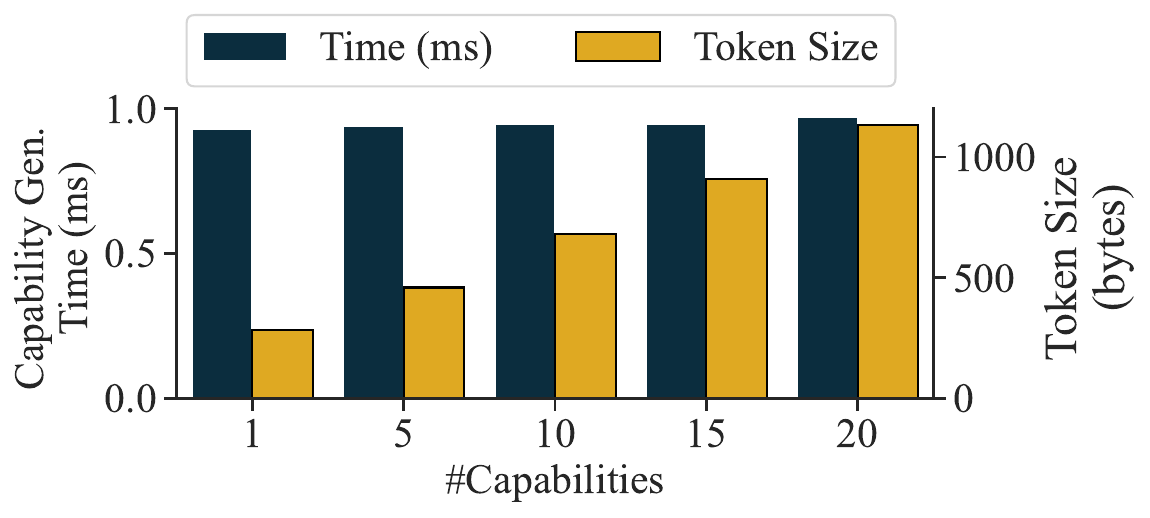}
    \caption{Token generation time and size for different number of capabilities.}
    \label{fig:varying_capabilities}
\end{figure}

\begin{figure*}[t]
    \centering
    \includegraphics[width=0.6\linewidth]{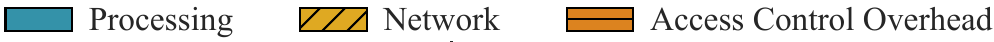}\\
    
    \subfloat[\centering Get Status]{{
        \includegraphics[width=0.3\linewidth]{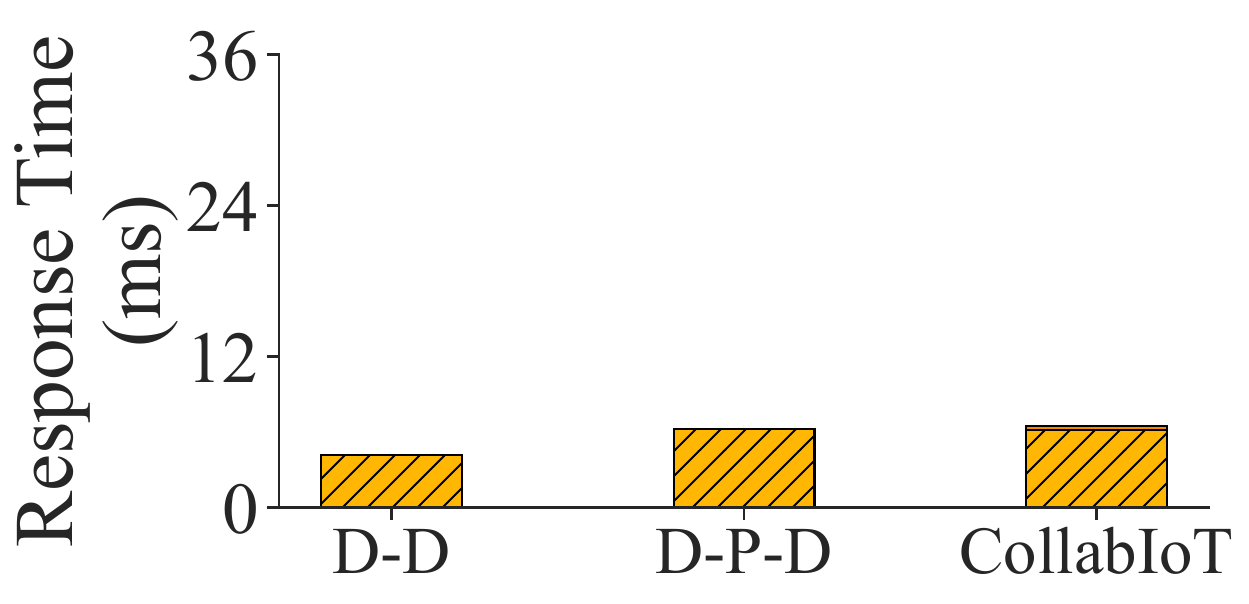}}
        \label{fig:proxy_overhead_ping}
    }%
    \quad
    \subfloat[\centering Load Image]{{
        \includegraphics[width=0.3\linewidth]{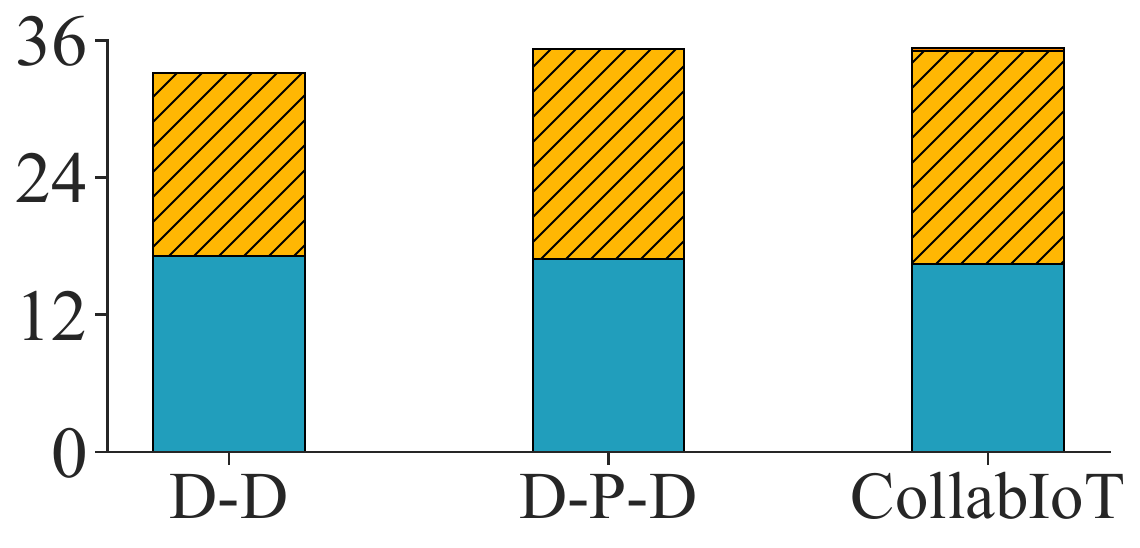}}
        \label{fig:proxy_overhead_IO}
    }%
    \quad
    \subfloat[\centering AI Processing]{{
        \includegraphics[width=0.3\linewidth]{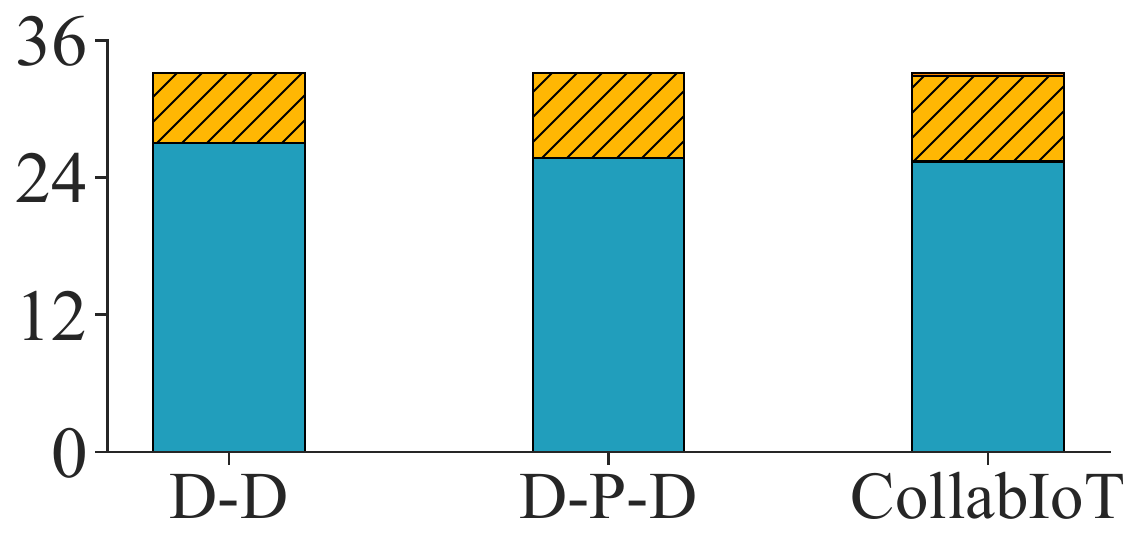}}
        \label{fig:proxy_overhead_com}
    }%
    \caption{Evaluating the performance of proxy-based architecture across three service types.}
    \label{fig:proxy_overhead}
\end{figure*}

\subsubsection{\textbf{Access Tokens}} 
~\autoref{fig:varying_capabilities} evaluates the performance of our capability-based authorization scheme across a different number of capabilities. The figure focuses on the capability generation time (ms) and the final token sizes (bytes). As shown, the capability generation time is constant across different numbers of capabilities ($\sim$1ms). The token size, however, increases sublinearly with the number of assigned capabilities, where the token size increases by 851B when the token contains 20 capabilities. 

\subsubsection{\textbf{Proxy Overhead}}
In this section, we evaluate the overheads of our lightweight proxy-based architecture compared to direct device-to-device interactions.
~\autoref{fig:proxy_overhead} evaluates the performance of proxy-based architecture in three service types: \texttt{get\_status, load\_image}, and \texttt{process\_image} services. 
~\autoref{fig:proxy_overhead_ping}) depicts the performance of \texttt{get\_status} request, \autoref{fig:proxy_overhead_IO}) implements an image load service (e.g., loading a camera feed), where the network overheads become more pronounced, and ~\autoref{fig:proxy_overhead_com} shows the time for the AI processing (e.g., image classification using Resnet-152), where most of the time is spent at the devices. We note that in all cases, the proxy adds up to 2 ms to the end-to-end response time, and access control overhead (including the token verification and enforcing rate limits) adds a maximum of 0.3 ms.

\begin{figure}[t]
    \centering
    \includegraphics[width=0.9\linewidth]{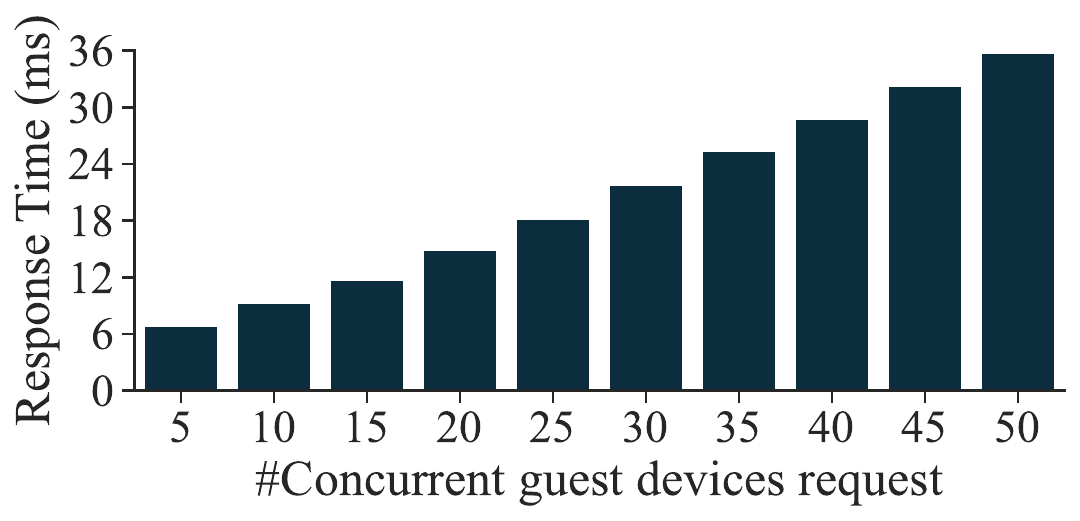}
    \caption{Response time of get status service for different concurrent guests.}
    \label{fig:concurrent_requests}
\end{figure}

\subsubsection{\textbf{Run-time Scalability}}

Next, we evaluate the scalability of our run-time environment.
~\autoref{fig:concurrent_requests} shows the end-to-end response time of the get status functionality, which includes the time for token verification (i.e., access control overhead), network time, and processing time. In this figure, we change the number of guests from 5 to 50 concurrent guest devices accessing a single device proxy. 
As shown in ~\autoref{fig:concurrent_requests}, the results demonstrate that the proxy can support multiple clients with small overheads. For instance, the results show that while a single client requires 6.46 ms per request (see \autoref{fig:proxy_overhead_ping}), five clients have an average response time of 6.83 ms, representing a 5.7\% overhead. Moreover, the results show that our system can withstand 50 clients with minimal overhead, resulting in a latency of 35.8 ms.

\begin{tcolorbox}[enhanced,size=fbox,drop shadow southwest,sharp corners, colback=white]
\noindent\textit{\textbf{Key Takeaways.} 
Our large-scale evaluation demonstrates the scalability of \systemName across different levels of policy complexities, including different attributes and capabilities. Additionally, our evaluation reveals that our proxy-based data plane incurs network overheads of up to 2 ms and access control overhead up to 0.3 ms. 
}
\end{tcolorbox}

\section{Related Work}
\label{sec:related-work}
\noindent\textbf{Automatic Policy Generation.} 
Policy generation frameworks have evolved from structured rule-based interfaces to machine learning–assisted generation, and more recently, to large language models (LLMs)-based generation. 
Earlier research enabled users to specify access policies using templates or constrained logic~\cite{damianou2001ponder,Rei,1333033,10.1145/3264905,10.1145/2938559.2938578}. While effective for simple rules, these approaches struggled with scalability and user comprehension as policy complexity increased. To overcome these limitations, subsequent work employed machine learning (ML) and natural language processing (NLP) to automate policy generation~\cite{alohaly2019automated, del2022systematic, heaps2021access, jayasundara2024sok, xia2022automated}, though such methods often required domain-specific retraining. More recent efforts leverage LLMs' ability to understand and generate structured outputs to directly generate policies from natural language~\cite{subramaniam2024intent, vatsa2025synthesizing, cheng2025:SayWhatYouMean, yao2024survey, sonune2025lmn}, thereby improving generalization and adaptability. However, these approaches lack mechanisms to validate the correctness of generated policies. In contrast, \systemName combines LLM-based synthesis with multi-stage validation to produce accurate, fine-grained, and enforceable access control policies aligned with user intent.

\noindent\textbf{Access Control in Transient Environments.}
Although central authorization approaches (e.g., 
OAuth 2.0~\cite{10.17487/RFC6749}) is widely used for delegating access via bearer tokens that get validated in a centralized authorization server. In decentralized settings, however, validating resources via a central authority may not be adequate, as it introduces unnecessary overhead. To address this issue, Capability-Based Access Control (CapBAC) offers an alternative for such settings by encoding access rights directly in signed tokens, allowing users to generate fine-grained and self-validating access tokens~\cite{gusmeroli2013capability, Fotiou2021:OAuthandVC, 9833873, 9709087}. 
While CapBAC approaches address fine-grained and centralization issues, they often lack mechanisms to configure access to devices automatically. Thus, Attribute-Based Access Control (ABAC) has been proposed to handle fine-grained, context-aware access in smart homes~\cite{Sikder2020:KRATOS,sikder2022s,goyal2022securing,ameer2022attribute,10.1145/3180457.3180464}, supporting multi-user and multi-device scenarios. 
\systemName draws parallels from hybrid models that combine ABAC and CapBAC~\cite{bertin2019access,10763520,9119088} and issue capability tokens based on attribute evaluation. Moreover, \systemName utilizes an LLM-based approach to facilitate policy generation.

\noindent\textbf{Seamless Collaboration in heterogeneous environments.} 
Many earlier efforts have addressed communication issues between heterogeneous devices.
Jini (Apache River)~\cite{Jini}, introduced mobile Java proxies to enable vendor-independent communication. However, its reliance on the Java Virtual Machine made it impractical for today’s resource-constrained IoT devices. \systemName adopts the proxy concept but improves it by localizing translation logic at the device edge, enabling lightweight RPCs from any client. Cloud-based platforms like IFTTT~\cite{IFTTT}, TTEO~\cite{yun2016tteo}, and Alexa Routines~\cite{AlexaRoutines} support user-defined automation via trigger-action rules but treat devices as passive and rely on centralized clouds, introducing latency and limiting real-time collaboration. \systemName, in contrast, enables direct, local communication and fine-grained access control. Decentralized systems like MOSDEN~\cite{6758734}, MATT~\cite{10.1145/3369823}, Coulson’s contact-action framework~\cite{7158188}, and SIoT~\cite{atzori2012social} explore opportunistic collaboration but lack runtime policy enforcement and vendor abstraction, limiting their use in dynamic settings. Standardization efforts like AllJoyn~\cite{Alljoyn}, IoTivity~\cite{IoTivity}, and Matter~\cite{Matter} aim for cross-vendor interoperability but require complex onboarding or firmware compliance. \systemName complements these efforts by supporting transient, heterogeneous devices through dynamic proxies that enforce access policies locally and abstract away vendor-specific APIs.

\section{Conclusion}
\label{sec:conclusion}
This paper presented \systemName, a system that enables secure and seamless device collaboration in transient IoT environments. \systemName address the security concerns by allowing users to define fine-grained access control policies and employs an LLM-based policy generation pipeline to generate validated access control policies from natural language descriptions.  To support real-time collaboration, \systemName automates authorization by distributing capability tokens to transient devices and employs lightweight proxies for policy enforcement. These proxies expose hardware-independent APIs, enabling interactions across heterogeneous devices.  We demonstrate that our LLM-based policy generation achieves 100\% accuracy in generating functional and correct policies. At runtime, our evaluation shows that our system configures new devices in $\sim$150 ms, and our proxy-based data plane incurs network overheads of up to 2 ms and access control overheads up to 0.3 ms. In future work, we plan to integrate \systemName with vendor-agnostic IoT frameworks such as Matter and extend the role of LLMs in device discovery and tagging.

\bibliographystyle{unsrt}
\bibliography{paper}

\end{document}